\documentclass[aps,pra,twocolumn,showpacs,superscriptaddress]{revtex4-1}

\usepackage{graphicx}
\usepackage{dcolumn}
\usepackage{amsmath}
\usepackage{epsfig}
\usepackage{bm}
\usepackage{amssymb}
\usepackage{natbib}
\usepackage{color}
\usepackage{hyperref}
\hypersetup{colorlinks=true,linkcolor=blue,citecolor=blue,urlcolor=black}
\usepackage{xspace}
\usepackage{array}
\usepackage{enumitem}
\usepackage{color}

\usepackage[squaren]{SIunits}

\newcommand{\ket}[1]{\ensuremath{{\left|#1\right\rangle}}\xspace}

\newcommand{\moy}[1]{\ensuremath{\langle#1\rangle}\xspace}
\newcommand{\braket}[2]{\ensuremath{\langle#1|#2\rangle}\xspace}

\newcommand{\modified}[1]{\textcolor{black}{#1}}

\newcommand{\remodified}[1]{\textcolor{black}{#1}}

\begin{document}

\title{Towards quantum simulation with circular Rydberg atoms}

\author{T.L. Nguyen}
\author{J.M. Raimond}
\author{C. Sayrin}
\author{R. Corti\~nas}
\author{T. Cantat-Moltrecht}
\author{F. Assemat}
\author{I. Dotsenko}
\author{S.~Gleyzes}
\author{S. Haroche}
\affiliation{Laboratoire Kastler Brossel, Coll{\`e}ge de France, CNRS, ENS-Universit\'e PSL, UPMC-Sorbonne Universit{\'e}, 11, place Marcelin Berthelot, 75231 Paris Cedex 05, France}
\author{G. Roux}
\author{Th. Jolicoeur}
\affiliation{LPTMS, CNRS, Univ. Paris-Sud, Universit\'e Paris-Saclay, 91405 Orsay, France}
\author{M. Brune}
\email{brune@lkb.ens.fr}
\affiliation{Laboratoire Kastler Brossel, Coll{\`e}ge de France, CNRS, ENS-Universit\'e PSL, UPMC-Sorbonne Universit{\'e}, 11, place Marcelin Berthelot, 75231 Paris Cedex 05, France}

\date{\today}


\begin{abstract}
The main objective of quantum simulation is an in-depth understanding of many-body physics. It is important for fundamental issues (quantum phase transitions, transport, \ldots) and for the development of innovative materials. Analytic approaches to many-body systems are limited and the huge size of their Hilbert space makes numerical simulations on classical computers intractable. A quantum simulator avoids these limitations by transcribing the system of interest into another, with the same dynamics but with interaction parameters under control and with experimental access to all relevant observables. Quantum simulation of spin systems is being explored with trapped ions, neutral atoms and superconducting devices. We propose here a new paradigm for quantum simulation of spin-$1/2$ arrays providing unprecedented flexibility and allowing one to explore domains beyond the reach of other platforms. It is based on laser-trapped circular Rydberg atoms. Their long intrinsic lifetimes combined with the inhibition of their microwave spontaneous emission and their low sensitivity to collisions and photoionization make trapping lifetimes in the minute range realistic with state-of-the-art techniques. Ultra-cold defect-free circular atom chains can be prepared by a variant of the evaporative cooling method. \modified{This method also leads to the detection of arbitrary spin observables with single-site resolution}. The proposed simulator realizes an XXZ spin-$1/2$ Hamiltonian with nearest-neighbor couplings ranging from a few to tens of kiloHertz.  All the model parameters can be \modified{dynamically} tuned at will, making a large range of simulations accessible. The system evolution can be followed over times in the range of seconds, long enough to be relevant for ground-state adiabatic preparation and for the study of thermalization, disorder or Floquet time crystals. \modified{The proposed platform already presents unrivaled features for quantum simulation of regular spin chains. We discuss extensions towards more general quantum simulations of interacting spin systems with full control on individual interactions.}
\end{abstract}

\maketitle

\section{Introduction}
\label{sec:introduction}

Understanding strongly-coupled many-body quantum systems is a problem of paramount importance. They present fascinating properties, such as quantum phase transitions~\cite{Sachdev2007}, topological phases~\cite{Bernevig2013}, quantum magnetism~\cite{Schollwoeck2008}, quantum transport~\cite{Dittrich1998} or many-body localization~\cite{Nandkishore2015}. Exploring this complex physics is essential for fundamental issues, such as fractional quantum Hall states~\cite{Cage2012} or high-temperature superconductivity~\cite{Phillips2012}. It may also lead to solutions to high-energy physics problems such as relativistic quantum field theories~\cite{Cirac2010}. Finally, it bears the promise of applications based on materials with engineered properties. 

The quantum many-body problem is all the more challenging that explicit analytical solutions are only available in a limited set of cases. Solid state experiments have to face the lack of access to some relevant quantities (entanglement properties for instance). Brute-force numerical exact diagonalization techniques face the exponential growth of the Hilbert space. In the restricted set of problems without the so-called sign problem~\cite{MX_TROYERSIGN05} there are successful algorithms from the quantum Monte-Carlo family that allow for numerically exact solutions~\cite{MX_SUZUKIMC93}. However, many interesting physical problems are outside of this class. In one-dimensional physics problems, the DMRG algorithm~\cite{White1992,White1993,Schollwoeck2005} is very successful but requires specific entanglement properties.

The ideal tool to address many-body physics would be a `quantum simulator'~\cite{QI_FEYNMANQUCOMP82,QI_LLOYDSIMUL96,MX_NORISIUML14}, transcribing the dynamics of the system of interest into another one that is under complete experimental control. Its parameters can be tuned nearly at will, all its observables can be measured. In principle, a general purpose quantum computer could be turned into a `digital' quantum simulator at the expense of an embarrassingly high amount of resources~\cite{QI_LLOYDSIMUL96,QI_SANDERSSIMCIRC12}. A more realistic approach is the `analog' quantum simulator~\cite{QI_MANOUSAKISANQSIM02}, with the same complexity (number of spins for instance) as the system of interest.  An analog simulator made up of a few tens of spins would already surpass any classical machine~\cite{QI_NORIQSREV09}.  Analog quantum simulation is one of the most promising domains of quantum information science. 

This paper proposes a new paradigm for analog quantum simulation of spin arrays, based on laser-trapped circular Rydberg atoms, protected from spontaneous emission decay~\cite{QC_KLEPPNERINHIB81} and reaching extremely long lifetimes in the minute range. It combines a deterministic preparation and read-out of defect-free chains containing a few tens of atoms.  The strong dipole-dipole interaction between the giant atomic dipoles emulates a fully tunable spin-$1/2$ XXZ chain Hamiltonian~\cite{Baxter1982}. The chain dynamics can be followed over one second for a chain containing a few tens of atoms, corresponding to $\approx10^5$ elementary exchange times. \modified{We show that available laser trapping techniques, using individual control of many laser traps~\cite{ArXiv_Bernien2017}, could even extend much further the realm of interest of this platform.} This analog simulator could supersede other platforms, even though they have already achieved impressive performance.

\subsection{State of the art}

Trapped ions~\cite{ION_WINELANDNOBEL13} are excellent tools for digital simulation~\cite{ION_BLATTMAPSIMUL13}, since they combine long coherence times, high-fidelity gates and individual unit-efficiency state-selective detection. The digital simulation of a QED process is a remarkable achievement~\cite{ION_BLATTQEDSIMUL16}. Ions are also well-suited for analog quantum simulation of spin arrays. The spin-spin interaction is simulated by a laser-induced coupling of the ions' internal states with their motional modes. This interaction can be tuned between a long-range regime (independent upon the distance between the ions) and a mid-range one (decreasing as the cube of the distance)~\cite{ION_MONROESPINCHAIN09,ION_MONROEMAGNETISM13}. Recent experiments demonstrated quantum random walks of excitations in spin-$1/2$ or spin-1 chains~\cite{ION_BLATTSPINCHAIN14,ION_MONROESPIN1SIMU15}, spectroscopy of spin waves~\cite{ION_BLATTSPINWAVE15}, many body localization~\cite{ION_MONROEMBL16} and thermalization~\cite{ION_SCHAETZTEHRM16}. First  2-D simulations of spin-squeezing with long-range interactions~\cite{ION_BOLLINGER2D16} have been reported. Engineered interactions in the nearest-neighbor regime of great interest are not available yet. 

Superconducting circuits are thriving, with qubits interacting directly or via their common coupling to cavities~\cite{QI_SCHEOLKOPFREVIEW13,QC_SCHEOLKOPFCAV04}. They are adapted to digital~\cite{QC_MARTINISDIGSIM15,QI_MARTINISSPINSIMUL16} or analog~\cite{MX_WALLRAFFMPS15,QC_MARTINISIMUL16} simulations. The experiments involved so far either only a few high-quality qubits~\cite{QC_WALLRAFFSPINSIMUL15,MX_MARTINISCHIRAL17}, a moderate number of damped systems~\cite{MX_HOUCKPHASE17} or even a large number of strongly damped ones~\cite{QI_DWAVEANNEAL14}, for which quantum speed-up is an open question~\cite{,QI_TROYERANNEAL15}.

Cold atoms in optical potentials are a remarkable platform for quantum simulation~\cite{MX_LEWENSTEINSIMU07,MX_BLOCHLMANYBODYREVIEW08,QI_DALIBARDCOLDATOMSIMULATOR12}. They can emulate the quantized conductance of a mesoscopic channel~\cite{MX_ESSLINGERQUNTCOND15}. Their joint coupling to an optical Fabry-Perot cavity implements the Dicke phase transition~\cite{MX_ESSLINGERDICKE10}, more perspectives being offered by photonic band-gap cavities~\cite{MX_KIMBLESIMULPROP14,QC_LUKINPHOTONGATE14,MX_KIMBLESPINCHAIN17}. Many experiments use optical lattices, with unit filling in the Mott-insulator regime~\cite{MX_HANSCHMOTT02} and individual site imaging~\cite{MX_MESCHEDETRANSPORT03,MX_BLOCHSITEIMAG10,MX_KUHRFERMIONIMAG15,MX_GREINERCORR16}. Inter-site tunneling and on-site interactions implement a Bose-Hubbard~\cite{MX_FERLAINOBH16} or Fermi-Hubbard~\cite{MX_ZWIERLEINSPINFERMICORR16} Hamiltonian, on which complex entanglement properties can be measured~\cite{MX_GREINERTHERMALIZE16,MX_GREINERCORR16}. Controlled disorder created by a speckle pattern~\cite{MX_LEWENSTIENSIMUREV10} leads to explorations of many-body localization~\cite{MX_BLOCHLOCALIZE15}. Experiments reach now domains beyond the grasp of theoretical methods and classical computations~\cite{MX_BLOCHMBL2D16}.  Lattice dynamical manipulations~\cite{MX_ARIMONDOLATTSHAKE07,QI_DALIBARDCOLDATOMSIMULATOR12} or multi-level atoms~\cite{MX_DALIBARDGAUGE10} open the way to the simulation of gauge fields and topological phases~\cite{MX_SPIELMANEDGESTATES15,MX_INGUSCIOEDGESTATE15,MX_GREINERLADDER17}. However, following long term dynamics, such as that of spin glasses, is challenging, since it requires very long lattice lifetimes. Alternative solutions with smaller lattice spacings and higher tunneling rates have been proposed~\cite{MX_CIRACSUPRALATTICE13,MX_KIMBLEATOMLATTICEPROP15} but not realized yet. Polar molecules~\cite{QI_DEMILLEPOLMOL02,MX_YECOLDMOLINTERAC13} or magnetic atoms~\cite{MX_PFAUDIP09,MX_FERLAINOBH16} can also be used to enhance the interactions.

Rydberg atoms~\cite{TXT_GALLAGHER} experience giant dipole-dipole interactions. The van der Waals potential~\cite{ENS_DENSEGAS} is in the MHz range for inter-atomic distances of a few microns. These interactions lead to the dipole blockade mechanism~\cite{QI_LUKINDIPOLEBLOCKADE01}: a resonant laser can excite only one Rydberg atom out of a micron-sized volume, since the first excited atom detunes all the others from laser resonance~\cite{MX_KUZMICHFULLBLOCKADE12,MX_BROWAYESBLOCKADE14}. This leads to non-classical excitation statistics~\cite{MX_WEIDEMULLERANTIBLOC10,MX_PILLETDISSRYD14,MX_SAFFMANATOMICFOCK14,MX_OTTSUPERATOM15,MX_PFAUAGGREGATES15,ENS_CHIPINTERACTION15}, to quantum gates~\cite{MX_SAFFMANRYDREVIEW10,MX_GRANGIERRYDENTANGLE10,MX_SAFFMANCNOT10,MX_BROWAEYSDD14}, to self-organization of Rydberg excitations~\cite{MX_BLOCHSTRUCTUREDRYDBERG12}, and to giant optical non-linearities~\cite{MX_GRANGIERNONLINRYD13,MX_ADAMSRYDGATEPROP14,MX_REMPERYDTRANS14,MX_LUKINBOUNDPHOTONS15,MX_HOFFERBERTHSINGLEPHOTRYD16,MX_VULETICPHOTONCOLL17}. These features are promising for quantum simulation~\cite{MX_LESANOVSKUYRYDCOMP11,MX_LESANOVSKYSPINCHAIN12,MX_WUSTERRYDTRANSP15}. Coherent excitation transport~\cite{MX_BROWAEYSSPINTRANSPORT15,MX_BROWAEYS2DRYD16} and synthetic spin arrays based on ground-state dressing with a Rydberg level~\cite{MX_BLOCHPSPINLAATICE16} have been demonstrated. However, the experiments have to face the finite lifetime of the laser-accessible Rydberg levels (few hundred of $\mu$s) and the blackbody-induced state transfers~\cite{MX_PORTORYDDRESS16}. Moreover, in all experiments so far, the Rydberg atoms are not trapped. The strong van der Waals forces between the atoms cause then a rapid explosion or collapse of the atomic ensemble~\cite{ENS_CHIPINTERACTION15}, limiting further its useful lifetime. Replacing the actual excitation to a Rydberg level by a ground-state laser dressing solves the problem only in part~\cite{MX_ZOLLERSPINRYD15}. Simulations of slow processes over long times are, for the time being, beyond the reach of low-angular-momentum Rydberg atom simulators.

\subsection{Principle of the proposed simulator}

\begin{figure}
\begin{center}
\includegraphics[width=7.5cm]{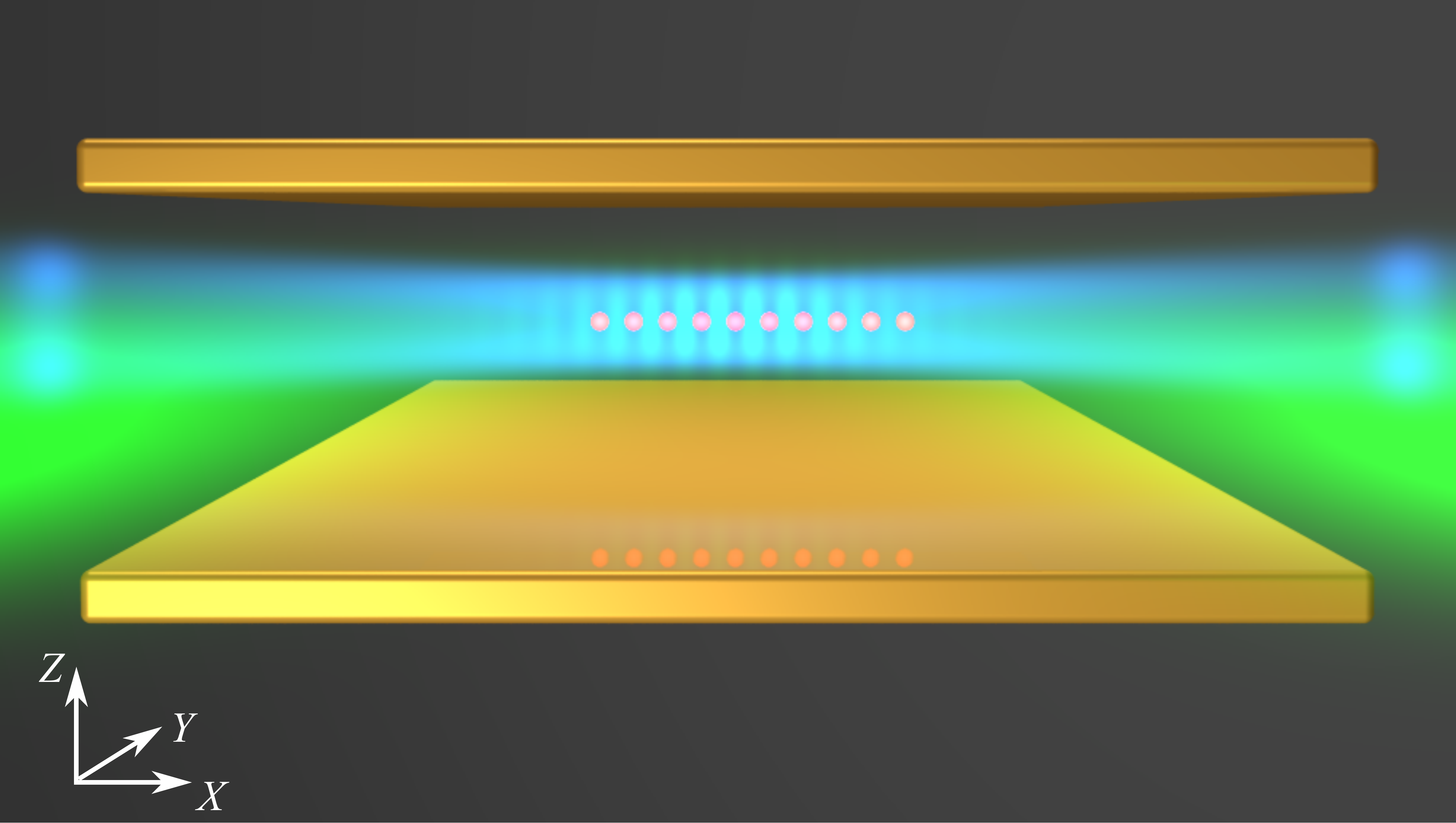}
\end{center}
\caption{Pictorial scheme of the proposed circular state quantum simulator.}
\label{fig:picscheme}
\end{figure}

We propose here a circular-state quantum simulator, schematized in Fig.~\ref{fig:picscheme}, which combines the best features of the other platforms and avoids some of their bottlenecks. Rydberg atoms in circular states, i.e., states with maximum angular momentum,  are trapped in the ponderomotive potential induced by laser fields~\cite{MX_FABRERYDHF76,MX_RAITHELTRAP00}. These low-field seekers are radially confined on the $OX$ axis (axis assignment in Fig.~\ref{fig:picscheme}) by a Laguerre-Gauss `hollow beam' at a 1~$\mu$m wavelength. They are longitudinally confined in a one-dimensional adjustable lattice produced by two 1~$\mu$m-wavelength beams, propagating in the $XOY$ plane at small angles with respect to the $OY$ axis. In the following, we will consider for the sake of definiteness two lattices with  inter-site spacings $d=5\ \mu$m and $d=7\ \mu$m, corresponding to a strong or moderate dipole-dipole interaction, respectively. The main decay channel of circular levels (spontaneous emission on the microwave transition towards the next lower circular level) is efficiently inhibited~\cite{QC_KLEPPNERINHIBITION85} by placing the atoms in a plane-parallel capacitor, which also provides a static electric field defining the quantization axis $OZ$ (the plane of the circular orbit is  thus parallel to the capacitor plates). A method based on a van der Waals variant of evaporative cooling~\cite{MX_KLEPPNEREVAP88} prepares deterministically long chains of atoms. \modified{It also leads to an efficient detection of individual atomic states with single-site resolution}. 

The spin-up and spin-down states of the simulator are encoded in the circular levels with principal quantum numbers 50 and 48, respectively, connected by a two-photon transition. The dipole-dipole interaction provides a general spin-$1/2$ XXZ chain Hamiltonian~\cite{Baxter1982} with nearest-neighbor interactions. Its parameters can be adjusted at will over a short time scale by tuning the static electric field and a near-resonant microwave dressing. This complete freedom in the choice of the model Hamiltonian is a unique feature of the circular state quantum simulator.

The dynamics of a chain with a few tens of spins can be followed over up to about $10^5$ spin-coupling times. The final state of each spin can be individually measured. Adiabatic evolutions through quantum phase transitions, sudden quenches and fast modulations of the interaction parameters are within reach. This proposal thus opens promising perspectives for the simulation of spin systems in a thermodynamically relevant limit, beyond the grasp of classical computing methods.

In Section~\ref{sec:vdw}, we recall the main properties of circular Rydberg atoms and discuss their dipole-dipole interaction. Additional details are given in Appendix~\ref{app:vdW}.  Section~\ref{sec:hamiltonian} is devoted to the  interaction Hamiltonian of an atom chain and to the rich phase diagram of the corresponding spin system, with details on the associated numerical simulations in Appendix~\ref{app:numerics}.  Section~\ref{sec:trap} is devoted to the laser trapping of circular atoms and to their protection from loss mechanisms, with technical details in Appendices~\ref{app:losses} and \ref{app:trap}. Section~\ref{sec:chain} is devoted to the deterministic preparation of a Rydberg atom lattice with unit filling (see also Appendix~\ref{app:evap}). Section~\ref{sec:simus} presents the results of state-of-the-art numerical simulations showing that the simulator reaches a thermodynamically relevant regime. We examine the most interesting perspectives in the concluding Section~\ref{sec:conclusion}.

\section{Circular Rydberg atoms and van~derWaals interaction}
\label{sec:vdw}

The circular states $\ket{nC}$ have a large principal quantum number $n$ and maximum orbital and magnetic quantum numbers: $\ell=|m|=n-1$~\cite{TXT_GALLAGHER}. They are the states closest to the circular orbit of the Bohr model, with a radius $r_n=a_0n^2$ ($a_0$: Bohr radius). Their wavefunction is a torus, with a small radius $r_n/\sqrt n$, centered on this orbit. This anisotropic orbit is stable only in a directing electric field  $\mathbf{F}$, normal to the orbit, defining the quantization axis $OZ$ and isolating the circular state from the hydrogenic manifold~\cite{ENS_DAHU} (Appendix~\ref{app:vdW}). The circular states cannot be excited directly from the ground state. Their preparation relies on a complex but efficient and fast process, combining laser and radio-frequency photons absorption~\cite{ENS_QZDEXP14}. These states have long radiative lifetimes, scaling as $n^5$ (25~ms for $\ket{48C}$). The microwave transitions between neighboring circular states are strongly coupled to the electromagnetic field. These remarkable properties make them ideal tools for experiments on fundamental quantum processes in cavity quantum electrodynamics experiments~\cite{ENS_OUP06,ENS_NOBELLECTURE13}.  

The large dipole matrix elements between circular levels make them particularly sensitive to the dipole-dipole interaction. Two atoms in the same circular state $\ket{nC}$ experience a van der Waals, second-order interaction proportional to $1/d^6$ ($d$: interatomic distance), repulsive in the proposed geometry (the interatomic axis, $OX$, is perpendicular to the quantization axis $OZ$, see Fig.~\ref{fig:picscheme}). For atoms in different circular states, $\ket{nC}$ and $\ket{pC}$, this interaction competes with the resonant F\"orster-like transfer of energy (`spin exchange') from one atom to the other: $\ket{nC,pC}\leftrightarrow\ket{pC,nC}$. This exchange process is at first order in the dipole-dipole interaction when $p=n\pm 1$. Scaling as $1/d^3$, it then overwhelms the repulsive interaction, realizing a spin model in which the spin exchange is by far the dominant interaction. With $p=n\pm 3$, the exchange is negligible compared to the van der Waals interaction. We chose here a more flexible simulator. With $p=n\pm 2$, the van der Waals and exchange interactions are of the same order of magnitude, scaling both as $1/d^6$. Their competition opens, as we show below, a wide range of possibilities to engineer interatomic potentials.

The dipole-dipole interaction mixes the circular states with neighboring elliptical states (Appendix~\ref{app:vdW}), since it breaks the cylindrical symmetry of the Stark effect.  These elliptical states have decay channels that are not inhibited by the capacitor (Appendix~\ref{app:losses}). This deleterious mixing effect can be reduced by using a large enough directing electric field $\mathbf{F}$ and a magnetic field $\mathbf{B}$ parallel to it. 

A careful optimization led us to choose the $\ket{50C}$ and $\ket{48C}$ states to represent the `spin-up' and `spin-down' states. With the field values $B=13$~Gauss and $6<F<12$ V/cm, the intrinsic lifetime of interacting atoms exceeds 90~s  for the smallest $d=5\ \mu$m interatomic distance. Lower principal quantum numbers would lead to an annoyingly small inhibition capacitor spacing. Higher principal quantum numbers would lead to larger spacings and dipole-dipole couplings. However, the transition frequencies between adjacent Rydberg manifolds is reduced and the lifetime reduction due to increased blackbody-induced transfer rates  (Appendix~\ref{app:losses}) is not compensated by the increase in couplings.

\begin{figure}
\begin{center}
\includegraphics[width=7.5cm]{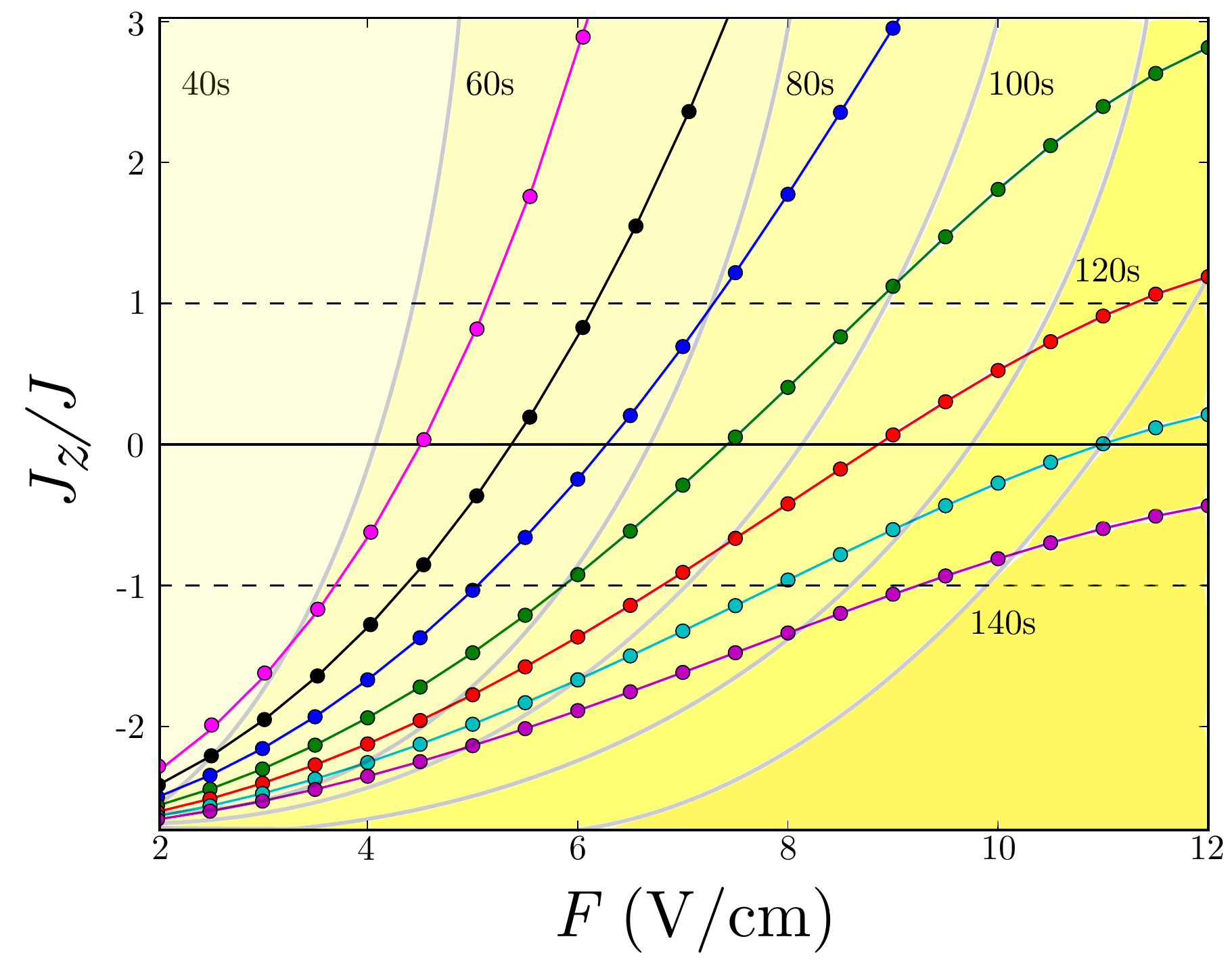}
\end{center}
\caption{Variation of $J_z$,  in units of $J$, with the directing electric field amplitude $F$. Dots result from the numerical diagonalization of the complete atomic Hamiltonian for $B=9,10,11,12,13, 14$ and $15$~Gauss (magenta, black, blue, green, red, cyan and purple dots respectively). The colored lines are a guide to the eye.  The horizontal solid line and the dotted lines correspond to the pure XY spin-$1/2$ exchange model and to the isotropic models, respectively. The shaded background and the light gray lines give a qualitative estimation of the lifetime of a pair of interacting atoms at a $d=5\ \mu$m distance. This estimation is based on explicit lifetime calculations for the $F$ and $B$ values corresponding to the plotted dots.}
\label{fig:jzj}
\end{figure}

The interaction Hamiltonian $V$ for a pair of atoms  reads, in terms of the atomic pseudo-spin operators (Appendix~\ref{app:vdW})
\begin{equation}
\frac{ V}{h} =\frac{\delta\zeta}{2}\left(\sigma^z_1 + \sigma^z_2\right) +J_z\,\sigma^z_1\sigma^z_2 +J\left(\sigma^x_1\sigma^x_2+\sigma^y_1\sigma^y_2\right) \ .
\end{equation}
The positive exchange term, $J$, is nearly independent of the directing electric field $F$. It is  proportional to $1/d^6$, strong ($17$~kHz) for $d=5\ \mu$m or moderate (2.3~kHz) for  $d=7\ \mu$m. The frequency shift $\delta\zeta$, of the order of $J$, also proportional to $1/d^6$, exhibits a slow field dependency (Appendix~\ref{app:vdW}). A unique feature of the circular state interaction is that $J_z$ varies significantly, from negative to positive values, with the electric field amplitude. The sign of $J_z$ can thus be controlled and the  ratio $J_z/J$ (independent on $d$) can be tuned over a large range by adjusting the control fields, as illustrated on Fig.~\ref{fig:jzj}. Over this complete range, the atomic lifetimes remain extremely long ($> 60$~s).

\section{The emulated XXZ model}
\label{sec:hamiltonian}

\subsection{Spin Chain Hamiltonian}

We now turn to a chain of $N$ interacting atoms at a constant spacing $d$. The Hamiltonian reads
\begin{align}
\label{Hint}
\frac{H_c}{h}=&\frac{\nu_0+\delta\zeta}{2}(\sigma^z_1 + \sigma^z_N)+\left(\frac{\nu_0}{2}+\delta\zeta \right)\sum_{j=2}^{N-1} \sigma^z_j \nonumber\\
&+\sum_{j=1}^{N-1}\left[ J_z \sigma^z_j\sigma^z_{j+1} +J(\sigma^x_j\sigma^x_{j+1}+\sigma^y_j\sigma^y_{j+1})\right] \,,
\end{align}
where $h\nu_0$ is the atomic transition energy ($\nu_0\approx2\times 55.97$~GHz for the $\ket{48C}\rightarrow\ket{50C}$ two-photon transition). We have here assumed that the pairwise dipole-dipole interactions are additive and we have neglected the next-nearest-neighbor interaction (64 times smaller than the nearest-neighbor one). Note that the atoms at the ends of the chain ($j=1$ and $j=N$) have a single neighbor and thus an energy shift ($h\delta\zeta$), which is half that of the atoms in the bulk ($j=2,\ldots, N-1$). The generalization of this Hamiltonian to arrays with higher dimensions is straightforward.

In this Hamiltonian, the atomic frequency is, by many orders of magnitude, the largest, making the ground state and the dynamics trivial. The situation is more interesting when driving the atoms by a $\sigma^+$-polarized classical field at a frequency $\nu$, close to resonance with the atomic two-photon transition ($\nu\simeq\nu_0/2$). The interaction with this field is, within an irrelevant phase choice for the classical driving field, represented by the effective two-level Hamiltonian
\begin{equation}
\frac{H_{d}}{h}=\Omega \cos(4\pi\nu t)\sum_{j=1}^N \sigma^x_j\ ,
\end{equation}
where $\Omega$ (considered as positive without loss of generality) is the effective Rabi frequency on the two-photon transition. Adding this term to the chain Hamiltonian, switching to an interaction representation defined by the unitary operator $U=\exp[i4\pi \nu t \sum_{j=1}^N (\sigma^z_j/2)]$ and using the rotating wave approximation, we get the final dressed-chain Hamiltonian
\begin{align}
\frac{{H} }{h}=&\frac{\Delta'}{2}(\sigma^z_1 + \sigma^z_N)+\frac{\Delta}{2}\sum_{j=2}^{N-1} \sigma^z_j +\frac{\Omega}{2}\sum_{j=1}^N \sigma^x_j  \nonumber\\
&+\sum_{j=1}^{N-1}\left[ J_z \sigma^z_j\sigma^z_{j+1} +J(\sigma^x_j\sigma^x_{j+1}+\sigma^y_j\sigma^y_{j+1})\right]  \, ,
\label{eq:Hamiltonian}
\end{align}
where $\Delta=(\nu_0+2\delta\zeta)-2\nu$ and $\Delta'=(\nu_0+\delta\zeta)-2\nu$. We recognize here a spin-$1/2$ XXZ chain Hamiltonian~\cite{DesCloizeaux1966,Yang1966,Yang1966a,Yang1966b,Baxter1982}, in which $J_z$ and $J$ describe the Ising coupling and spin-flip exchange, respectively. The detuning $\Delta$ plays the role of an effective longitudinal magnetic field, while $\Omega$ is an effective transverse field.

The field-independent $J$ term defines the fundamental exchange time scale for this Hamiltonian,  $\tau_{ex}=1/(4J)=14.7\ \mu$s at $d=5\ \mu$m and 108$\ \mu$s at $d=7\ \mu$m.  A unique feature of the  simulator is that all other parameters of the Hamiltonian are under experimental control. The $\Delta$ and $\Omega$ parameters are determined by the classical microwave source dressing the atomic transition and $J_z$ is controlled by the directing fields $F$ and $B$ (Fig.~\ref{fig:jzj}).  All the Hamiltonian parameters can thus be changed or modulated over a nanosecond time scale, infinitely short as compared to $\tau_{ex}$. This is a unique feature of this simulator.

\subsection{Phase diagram}
\label{sec:simu}

\begin{figure}[t]
\centering
\includegraphics[width=0.85\columnwidth,clip]{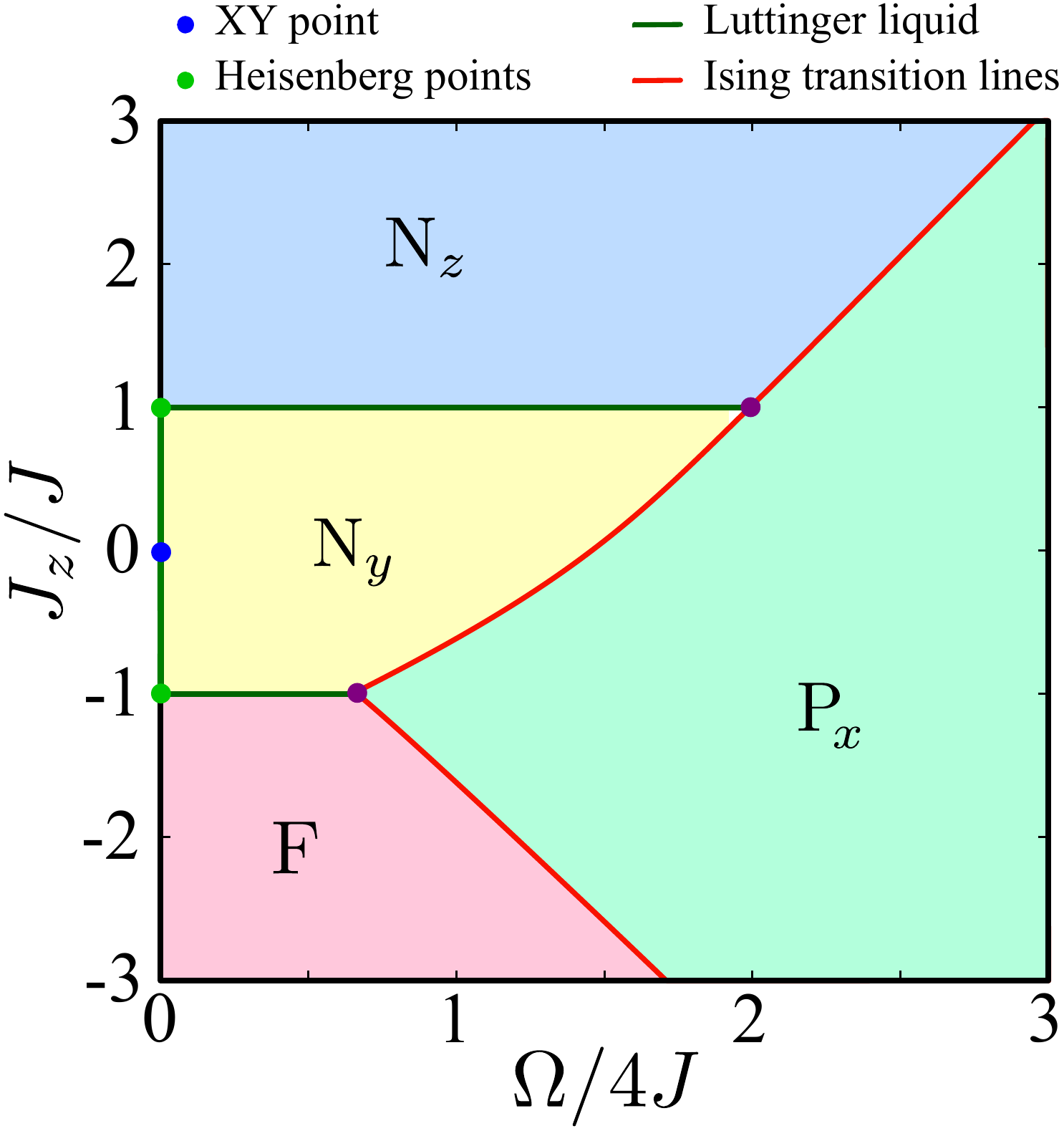}
\caption{Sketch of the phase diagram of Hamiltonian \eqref{eq:XXZ-omega} based on the results of Fig.~\ref{fig:numphasediag} and Ref.~\onlinecite{MX_DMITRIEV02}.}
\label{fig:phasediag}
\end{figure}

\begin{figure*}[t]
\centering
\includegraphics[width=\textwidth,clip]{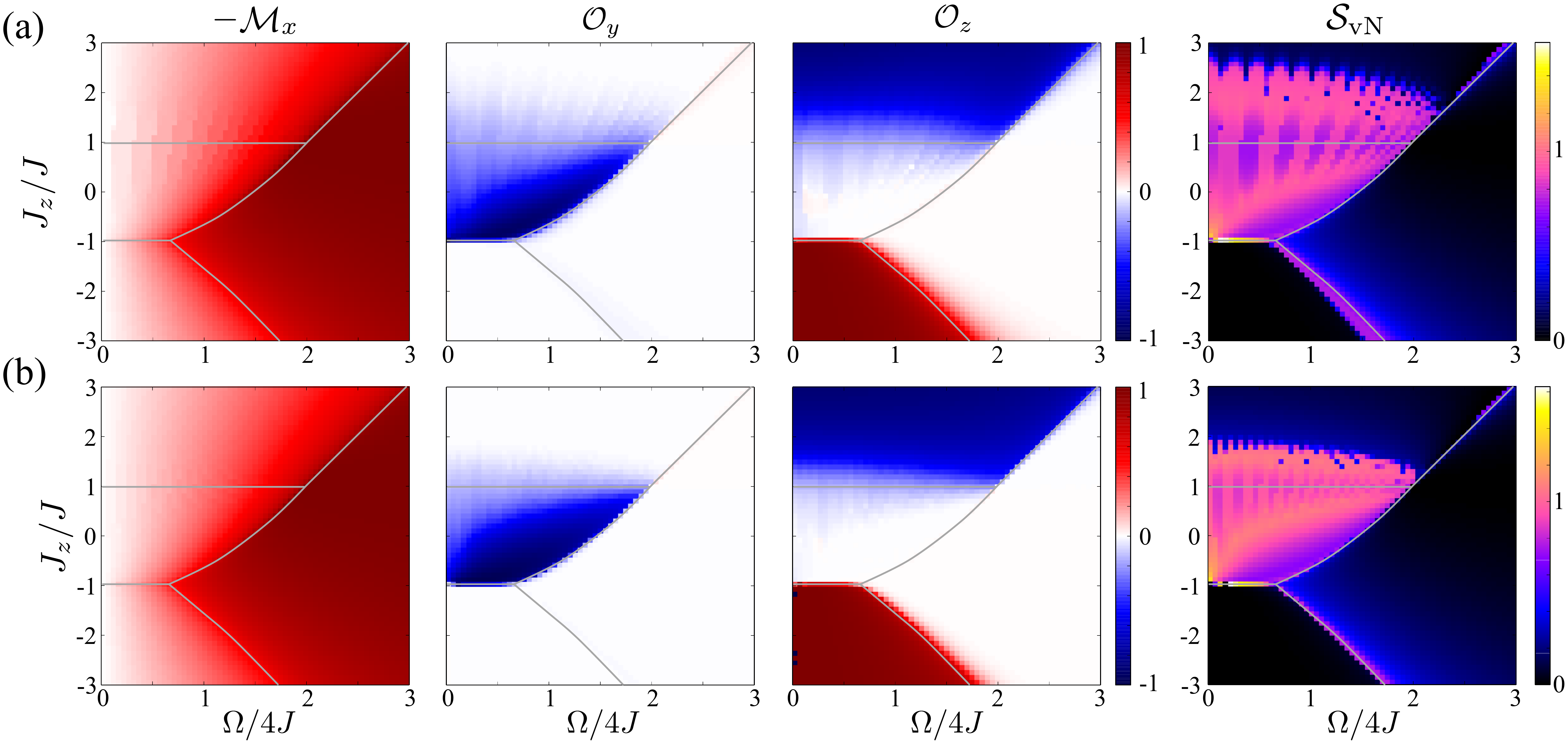}
\caption{\textbf{Numerical phase diagram of the XXZ in a transverse field}. {(a)}: MPS results for the order parameters $\mathcal{M}_x$, $\mathcal{O}_y$, $\mathcal{O}_z$ and von Neumann entropy $\mathcal{S}_\text{vN}$ (from left to right) for the Hamiltonian \eqref{eq:XXZ-omega} on an open chain with $N=40$ spins.{(b)}: same data for a  $N=90$ open spin chain. The order parameters $\mathcal{O}_{y,z}$ defined in \eqref{eq:ordercorrr} are computed with $r=17$ for $N=40$ and $r=31$ for $N=90$. Red regions represent ferromagnetic ordering while blue ones represent antiferromagnetic (N\'eel) ordering.  The gray lines are guides to the eyes for the quantum phase transition lines, inferred from symmetry arguments (horizontal lines) and from the von Neumann entropy plot for $N=90$. They have been used to delineate the phases in Fig.~\ref{fig:phasediag}}
\label{fig:numphasediag}
\end{figure*}

The  $\Delta=0$ case already provides a rich ground-state phase diagram, spanning a variety of key many-body problems.
In this Section, we review this diagram in the thermodynamic limit for the bulk of the simulator, putting aside the edge effects. Setting $\Delta=\Delta'=0$ in \eqref{eq:Hamiltonian}, the generic Hamiltonian reads
\begin{equation}
\label{eq:XXZ-omega}
\frac{H}{h} = \sum_j \Big[ J_z \sigma_j^z\sigma_{j+1}^z + J(\sigma_j^x\sigma_{j+1}^x + \sigma_j^y\sigma_{j+1}^y) +\frac{\Omega}{2}\sigma_j^x\Big]\,,
\end{equation}
which boils down to the XXZ model in a transverse field~\cite{Kurmann1982,Mueller1985,Mori1995,Hieida2001,MX_DMITRIEV02,MX_DMITRIEVXXZ02,Dutta2003}. This model is relevant, in particular, in the interesting physics of the Cs$_2$CoCl$_4$~\cite{Kenzelmann2002,Breunig2013} or BaCo$_2$V$_2$O$_8$~\cite{Grenier2015} quantum magnets.
Its phase diagram is sketched on Fig.~\ref{fig:phasediag} and exhibits interesting quantum phase transitions.

The main four phases are associated with different symmetry breakings of the generic Ising symmetries $\mathbb{Z}^y_2\otimes\mathbb{Z}^z_2$ ($\sigma_j^y\to -\sigma_j^y$ and $\sigma_j^z\to -\sigma_j^z$).
The competition between these phases is driven by the sign and strength of the $J_z/J$ parameter and by the magnitude of the transverse field $\Omega$. At large $\Omega/4J$, the field polarizes all spins close to the $x$-direction. This phase is gapped, does not break any symmetry and has a non-degenerate ground state. Using the terminology of the Ising model in a transverse field~\cite{Dutta2015}, we call it the ``paramagnetic phase'' (although it is ferromagnetically ordered along the $x$-direction) and denote it by P$_x$. This phase is separated from the others by Ising transition lines (red lines in Fig.~\ref{fig:phasediag}). 

The three symmetry-breaking phases stem from the $\Omega=0$ line corresponding to the  pure XXZ model. This model has three phases: a gapped ferromagnetic phase, F, for $J_z<-J$, a gapless (critical) Luttinger liquid phase~\cite{Haldane1981,Giamarchi2004} for $-J < J_z < J$ and a gapped N\'eel phase, N$_z$, along the $z$-direction, for $J_z>J$. The F and  N$_z$ phases  have doubly degenerate ground states and break the $\mathbb{Z}^z_2$ symmetry, with an additional breaking of translational symmetry for the N\'eel phase. When a transverse magnetic field is applied ($\Omega\not = 0$), the two gapped  F and  N$_z$  phases are stable until the gap closes at the Ising transition line, at which the system enters the P$_x$ phase. For the Luttinger liquid phase, a non-zero transverse field immediately opens a gap. The associated broken symmetry is $\mathbb{Z}^y_2$, corresponding to a N\'eel ordering in the $y$-direction (N$_y$ phase). This order is eventually destroyed by the transverse field through an Ising transition toward the P$_x$ phase.

The boundaries between the three phases, F,  N$_y$ and N$_z$ (green horizontal lines in Fig.~\ref{fig:phasediag}), with broken symmetries  emerge from the Heisenberg points $J_z = \pm J, \Omega=0$. Along these lines, the gapless system presents additional symmetries. Indeed,  the Heisenberg points  correspond to a SU(2) symmetry, which, under the application of the transverse field ($\Omega\not=0$), is reduced to U(1). The upper line $J_z=J$ corresponds to the Heisenberg model under an external field~\cite{Yang1966,Yang1966a}, for which a Luttinger liquid phase survives up to the critical field $\Omega_c/4J = 2$, at which a commensurate-incommensurate transition occurs~\cite{Pokrovsky1979,Schulz1980}.  On the opposite Heisenberg point $J_z = -J$, the transformation $\sigma_j^z\to (-1)^j\sigma_j^z$ maps the model onto the ferromagnetic Heisenberg chain. It has, as the other Heisenberg point, a SU(2) symmetry, lowered to U(1) when the transverse field is applied. Thus another straight critical line emerges from this Heisenberg point, separating N$_y$ from F. Due to the model mapping transformation, this coexistence line ends with a lower critical field than the $J_z=+J$ one~\cite{Alcaraz1995}.

This spin-$1/2$ model presents other remarkable features. The integrability of the model is an essential concept to discuss relaxation and thermalization. The model is integrable by the Bethe ansatz when $\Omega=0$ and on the critical lines emerging from the Heisenberg points. In particular, $\Omega = J_z = 0$ corresponds to the XY model that maps onto free fermions~\cite{Barouch1971}. In the $J=0$ limit, the model maps onto the (anti)ferromagnetic Ising model in a transverse field, which also maps onto free fermions~\cite{Pfeuty1970}, and is thus integrable. Away from these limits, the model is non-integrable.

The qualitative plot  of Fig.~\ref{fig:phasediag} is supported by numerical results based on matrix-product state (MPS) simulations~\cite{White1992,White1993,Schollwoeck2005,MX_SCHOLLWOEK11,ITensor} (Appendix~\ref{app:numerics}). We define the average magnetization along the axis $O\alpha$ ($\alpha=x,y,z$) as
\begin{equation}
\label{eq:magnetization}
\mathcal{M}_{\alpha} = \frac{1}{N}\sum_{j=1}^N \moy{\sigma_j^{\alpha}} \;.
\end{equation}
For symmetry reasons, $\mathcal{M}_{y,z}$ must be zero on non-degenerate finite-size ground state.
Therefore, the ordering of the spins is better captured by order parameters defined from correlations as
\begin{equation}
\label{eq:ordercorrr}
\mathcal{O}_{\alpha} =\text{sign}(C_{\alpha}) \sqrt{|C_{\alpha}|}\; \text{with}\; C_{\alpha} =  \moy{\sigma_j^{\alpha}\sigma_{j+r}^{\alpha}} \;,
\end{equation}
where $\alpha = y,z$, $j=N/2$, and where $r$ is ``large enough'', to be specified for a given $N$.

We plot in Fig.~\ref{fig:numphasediag} the magnetization and order parameters along the three spin axes as a function of $\Omega/4J$ and $J_z/J$ for $N=40$ and $N=90$ open spin chains. The first column shows that, as expected, the magnetization $\mathcal M_x$ increases steadily with $\Omega/4J$. The region with a large $\mathcal M_x$ value corresponds to the P$_x$ phase. Along the $J_z=J$ line and for $N=40$, we observe magnetization plateaus, corresponding to a succession of ground states with fixed total magnetization along $x$. These finite-size effects are gradually smoothed out away from this line~\cite{Mueller1985,MX_DMITRIEV02}.

The order parameters $\mathcal O_y$ and $\mathcal O_z$  show the strength of N\'eel and ferromagnetic ordering across the phase diagram. While most phase transitions are rather steep, the N$_y \leftrightarrow$ N$_z$ transition at $J_z=J$ is much smoother  due to strong finite-size effects. In this region, the gaps are indeed the smallest (the Luttinger liquid to N$_z$ transition is of the Berezinskii-Kosterlitz-Thouless type~\cite{Berezinskii1971,Kosterlitz1973,Kosterlitz1974}). 

The features of the phase diagram and its finite-size effects are also conspicuous when plotting the von Neumann entropy $S_\text{vN} = -\text{Tr}[\rho \ln \rho]$ where $\rho$ is the reduced density-matrix of the first $N/2$ spins in the chain. Along the critical lines, one expects~\cite{Calabrese2004,Calabrese2009} a logarithmic divergence of the entropy $S_{\text{vN}} \simeq \frac{c}{6} \ln N$ (for open boundary conditions) with $c=1$ for Luttinger liquid phases and $c=1/2$ for Ising transitions. In the gapped phases, the entropy remains finite, and decreases when the gap increases. It displays plateaus along the $J_z=J$ line reminiscent of the magnetization plateaus. The rapid variation of the entropy when increasing $J_z/J$ within the N$_z$ phase is due to fact that the MPS variational state breaks the $\mathbb{Z}^z_2$ symmetry (see Appendix~\ref{app:numerics}). 

Figure~\ref{fig:numphasediag} shows that the chain Hamiltonian exhibits a wide variety of interesting behaviors. It also shows that, in most regions, finite size effects are not too large, since a good approximation of the thermodynamical limit can be reached with 40 atoms only. \modified{The observation of this phase diagram would be an excellent benchmark for the operation of the simulator. It would make one confident that the results of more challenging dynamical experiments could be trusted, even in domains where direct calculations are not available and where interesting questions are still opened.}

\section{Preservation and trapping of circular Rydberg atoms}
\label{sec:trap}

\subsection{Circular atoms lifetime}

These remarkable features of the spin-chain Hamiltonian are only relevant if the circular atoms can be preserved and trapped for times much longer than $\tau_{ex}$, even much longer than their natural lifetime ($\Gamma_{48}^{-1}=25$~ms for $\ket{48C}$). They should thus be protected from spontaneous emission and from other loss mechanisms. We show in this Section that this ambitious goal can be achieved with state-of-the-art techniques.

D. Kleppner pointed-out~\cite{QC_KLEPPNERINHIB81} and experimentally demonstrated~\cite{QC_KLEPPNERINHIBITION85} that spontaneous emission can be inhibited by placing atoms in a structure with no field mode close to resonance with the atomic transition. The unique spontaneous decay channel for the circular states in a zero temperature environment is a $\sigma^+$-polarized transition towards the next lower circular state. It is inhibited in the plane-parallel capacitor providing $\mathbf{F}$ when its plates are separated by a distance $D$ smaller than half the radiated wavelength, $\lambda= 4.9$~mm for the $\ket{48C}\rightarrow\ket{47C}$ transition. In an ideal, infinite capacitor, the inhibition is complete and the circular level lifetime is infinite. 

\begin{figure}
\begin{center}
\includegraphics[width=7.5cm]{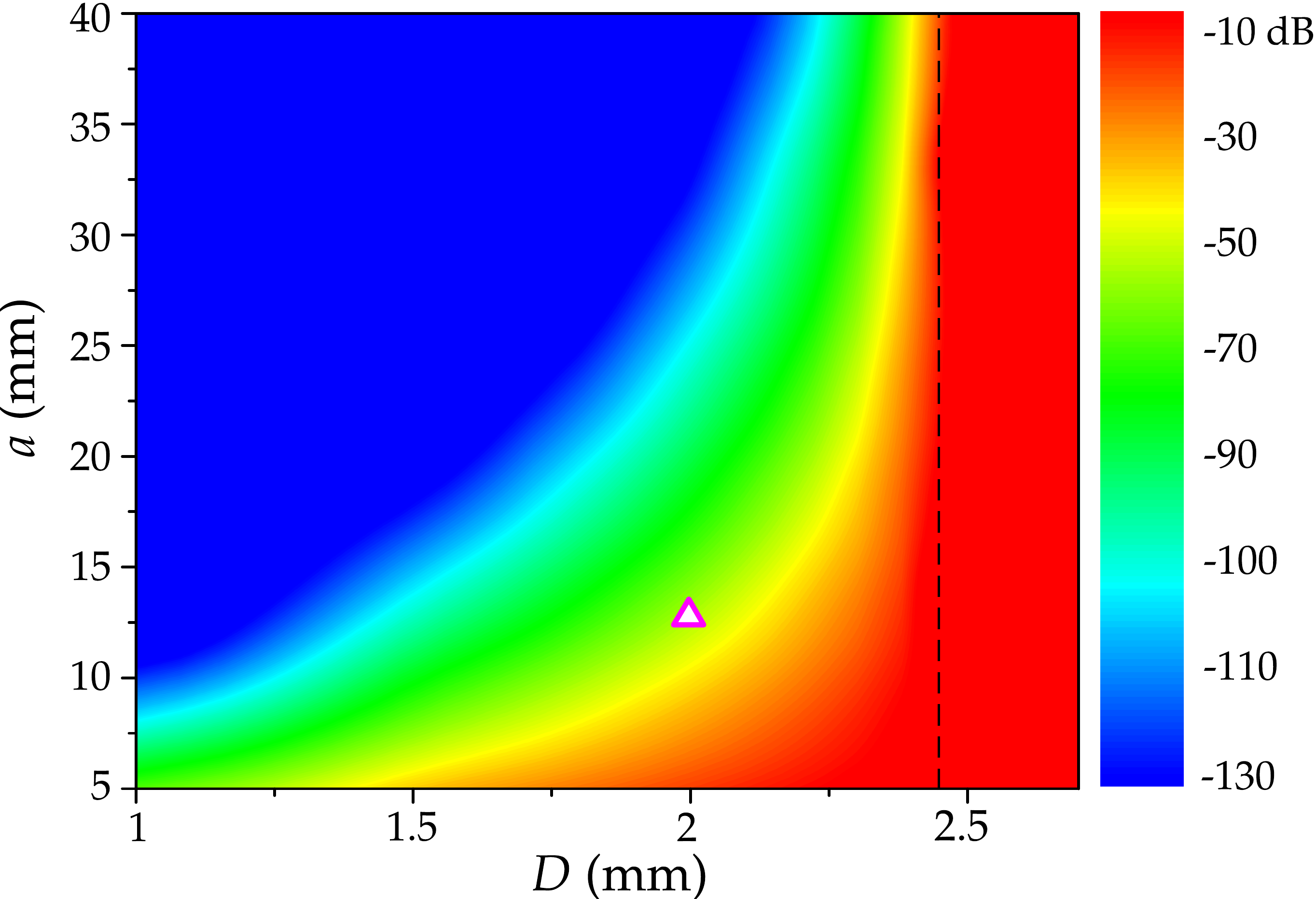}
\end{center} 
\caption{Spontaneous inhibition ratio, $\Gamma/\Gamma_{48}$ (log scale) as a function of the capacitor spacing $D$ and size $a$. The dashed vertical line corresponds to $D=\lambda/2$. The open red triangle shows the chosen operation point $D=2$~mm and $a=13$~mm with a 50~dB inhibition.}
\label{fig:inhib}
\end{figure}

A more realistic calculation should take into account the finite size and conductivity of the capacitor. We have numerically computed the residual spontaneous emission rate, $\Gamma$, for a capacitor with square plates (made up of gold cooled below 1~K) of side $a$, using the CST-studio software suite (Appendix~\ref{app:losses}). Figure~\ref{fig:inhib} shows the ratio $\Gamma/\Gamma_{48}$ as a function of $a$ and $D$. The inhibition is large as soon as  $a$ is larger than 10~mm. We choose for the following discussion an operating point with $D=2$~mm and $a=13$~mm, corresponding to a 50~dB inhibition rate, i.e., to a $\simeq 2500$~s lifetime for $\ket{48C}$. Note that the spontaneous emission inhibition for $\ket{50C}$ is even stronger, since the emission wavelength is larger. The opening between the capacitor plates is large enough to provide convenient optical access to the trapping region. 

The capacitor also inhibits the $\sigma^+$-polarized dressing microwave required, in particular, to engineer the chain Hamiltonian $H$. However, due to the sensitivity of Rydberg atoms to microwave fields, this drive requires only a low power. It can thus be applied on the atoms in an evanescent mode to which a powerful enough source is coupled, for instance through tiny ($<0.15$~mm diameter) irises pierced in the capacitor plates. According to simulations, these irises do not significantly affect the spontaneous emission inhibition.

Spurious effects conspire to reduce the lifetime (Appendix~\ref{app:losses}). Blackbody photons induce a $\pi$-polarized transition from the circular state towards elliptical states in a higher manifold. The transition rate for this polarization is enhanced by a factor $\simeq 2$ in the capacitor. Cryogenic temperatures are thus required to limit this effect. We assume $T=0.4$~K, a typical base temperature for $^3$He refrigerators. The effect of the collisions with the background gas is small for  a background pressure in the $10^{-14}$ torr range, accessible in a cryogenic environment~\cite{MX_GABRIELSEANTIPROTON90,ION_WERTHCRYOION98}. We must also include in the loss mechanisms the contamination by elliptical states due to the dipole-dipole interaction, photoionization, which turns out to be quite negligible for circular states, or the elastic diffusion of trapping-lasers photons. 

We finally find (Appendix~\ref{app:losses}) that  the levels lifetimes, including all foreseeable loss mechanisms, exceeds 53~s in the useful range of $F$ values (even longer lifetimes can be reached by increasing further $F$ and $B$ (at the expense of a reduced tunability of $J_z$ in the latter case). A  40-atom chain is thus expected to have a useful lifetime of at least 1.1~s, corresponding to  $8\times10^4$ spin exchange periods $\tau_{ex}$ at $d=5\ \mu$m. That one  can follow the dynamics of a spin chain over such long times is a unique feature of the circular state quantum simulator.

\subsection{Circular atom trapping}

The circular atoms must obviously be trapped in order to take benefit of these long lifetimes. Trapping them through the Stark or Zeeman effects has been proposed~\cite{ENS_RYDBERGTRAP04,ENS_EPJDTRAP} or realized~\cite{MX_RAITHELCIRCULARTRAP13}. These techniques, however, do not lead to flexible trap architectures.  We consider instead, following~\cite{MX_RAITHELTRAP00}, an optical laser trap.

The nearly free valence electron of the circular atom experiences a positive ponderomotive energy~\cite{MX_FABRERYDHF76} proportional to the laser intensity~$I$,
\begin{equation}
{\cal E}=\frac{e^2}{2m_e\,\varepsilon_0 c\,\omega_L^2} I\ ,
\end{equation}
where $e$ and $m_e$ are the electron's charge and mass, respectively, and where $\omega_L$ is the laser angular frequency (much larger than the electron's orbital frequency). The electron is thus attracted towards low intensity regions. The ponderomotive energy is 14.8~MHz (about 1~mK) in the 10~$\mu$m waist of a 1~W, 1~$\mu$m-wavelength laser. It is about ten times larger than the potential experienced by a ground-state rubidium atom in the same conditions.

The electronic attraction towards intensity minima is transmitted to the Rydberg atom as a whole (note that the ponderomotive energy of the ionic core is quite negligible due to its large mass). We propose to radially trap the rubidium atoms along the $OX$ axis (Fig.~\ref{fig:picscheme}) by a 0.5~W, 1~$\mu$m-wavelength hollow beam in a (0,1) Laguerre-Gauss (LG) mode focused to a 7~$\mu$m waist~\cite{MX_PRUVOSTCHANNEL14}. The transverse trapping frequencies are then $\omega_{Y}=\omega_Z=2\pi\times\,12$~kHz. At the edges of the inhibition capacitor, the LG beam diameter is 0.6~mm. The laser power hitting the plates (60~nW)  and dissipated in the cold environment is thus much less than the cooling power of the $^3$He refrigerator. 

\begin{figure}
\begin{center}
\includegraphics[width=7cm]{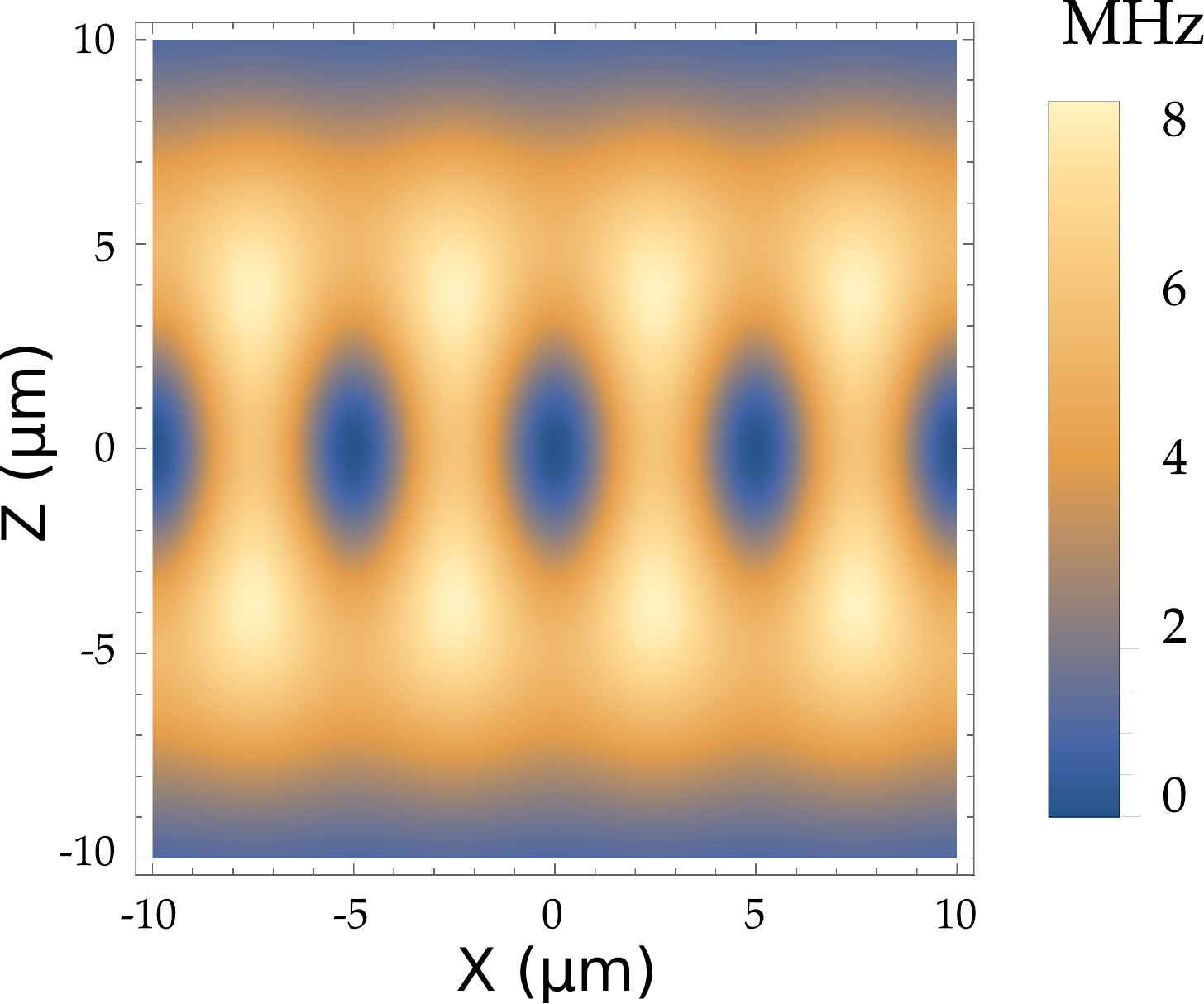}
\end{center} 
\caption{Cut in the $XOZ$ plane of the ponderomotive potential produced by a Laguerre-Gauss and two interfering gaussian beams at a  1~$\mu$m-wavelength. The potential values are given in frequency units by the color map on the right.}
\label{fig:trappot}
\end{figure}

The longitudinal lattice (along $OX$) should have an inter-site spacing adjustable at least between 5 and  7$\mu$m and provide a tight confinement  to reduce the variations of the dipole-dipole interactions due to the residual atomic motion. Note that the residual motion along the transverse axes is much less worrisome, acting only at the second order on the interatomic distance. In order to get a simply adjustable spacing, we suggest to use the interference at a small angle between two 1~$\mu$m-wavelength laser Gaussian beams, offset in frequency by a few tens of MHz with respect to the LG beam to avoid interferences with the transverse trap beam. They propagate in the $XOY$ plane at an angle $\pm\theta$ with respect to the $OY$ axis. For $d=5\ \mu$m, $\theta= 5.7\degree$ ( $\theta=4.1\degree$ for $d=7\ \mu$m). Their waist is 7~$\mu$m along $OZ$ and 200~$\mu$m along $OX$, so as to cover the whole length of the chain. With a power of 1.45~W in each beam  for $d=5\ \mu$m (2.8W for  $d=7\ \mu$m), we get $\omega_X=2\pi\times\,24$~kHz and a longitudinal trap depth of nearly 4~MHz, i.e. 200~$\mu$K  (Appendix~\ref{app:trap}). The power hitting the capacitor is also negligible for these beams. Figure~\ref{fig:trappot} presents the total ponderomotive potential for $d=5\ \mu$m. The deep traps are regularly spaced along the $OX$ axis. The extent of the atomic motion ground-state in these nearly harmonic traps is $\Delta X_0=50$~nm.

Note that, for a position-dependent laser intensity, the potential acting on the atom is the average of the ponderomotive energy over the atomic orbital~\cite{MX_RAITHELTRAP00}.  We show (Appendix~\ref{app:trap}) that this effect plays no role when the atom remains in the harmonic region close to the bottom of the trap. We also estimate the decoherence due to the atomic motion in the residual trap anharmonicity. The coherence time ($\approx$0.2~s) corrresponds to  $ 10^4\tau_{ex}$ at $d=5\ \mu$m.

We have suggested here a set of operating parameters adapted to the conservation of a strongly interacting long chain over extended times. Other compromises can be made, according to the experimental goals. Smaller couplings can be obtained with a larger inter-site spacing $d$, limiting the impact of the residual atomic motion of the chain dynamics (Section~\ref{sec:simus}). Much tighter traps can be obtained with higher laser powers, at the expense of a reduced lifetime. Longer lifetimes can be reached in very high electric and magnetic fields, at the expense of a reduced tunability of the Hamiltonian parameters.

\section{Deterministic preparation and detection of circular atom chains}
\label{sec:chain}

The $N$-atom chain must be prepared deterministically. Techniques based on the Mott transition~\cite{MX_HANSCHMOTT02} achieve a unit filling of a ground-state atom lattice. They are not easily applicable to the large lattice spacings envisioned here. Real-time feedback allows one to prepare regular arrays of independent dipole traps with unit filling~\cite{MX_BROWAEYSARRAY16,MX_LUKINARRAY16}. However, the preparation of circular levels from the ground state has a finite efficiency, leading to gaps in the final Rydberg chain. The dipole blockade mechanism could lead to nearly regular Rydberg atoms arrangements after the excitation of a BEC or of a lattice~\cite{MX_LUKINBLOCKADECRYSTAL10,MX_BLOCHSTRUCTUREDRYDBERG12} but, according to our simulations~\cite{ENS_RYDEXCITEXX}, interatomic distance variations are  large and lead to an excessive atomic motion in the final traps. 

We thus discuss in this Section an innovative chain preparation method based on a variant of evaporative cooling~\cite{MX_KLEPPNEREVAP88}. Its principle is to start with an irregular chain and a large random number of atoms trapped in a laser tube and to progressively compress and `evaporate' this chain until the required interatomic spacing and atom number are reached. The evaporation provides cooling nearly down to the ground state of the trap, leading to very small motional effects and dephasing. We show that the chain evaporation technique also leads to an efficient state-selective individual detection of each atom.

\begin{figure}
\begin{center}
\includegraphics[width=7.5cm]{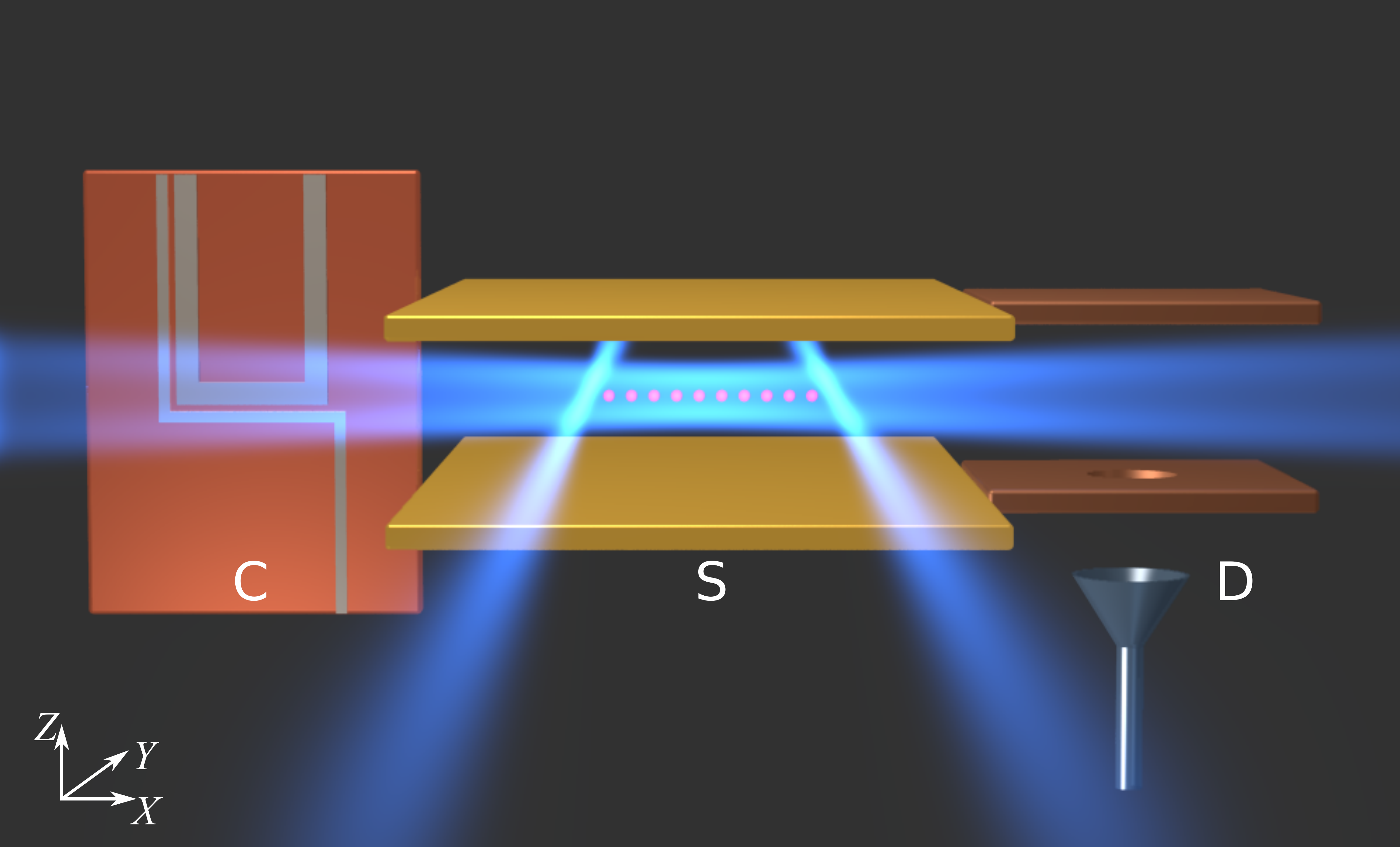}
\end{center} 
\caption{Sketch of the proposed chain preparation and detection sequence. $C$: atom chip. $S$: science capacitor. $D$: field-ionization detection region. Note the axes orientation in the lower left.}
\label{fig:compscheme}
\end{figure}

Figure~\ref{fig:compscheme} presents a conceptual scheme of the experiment. The sequence (detailed in Appendix~\ref{app:evap}) begins with the preparation near a superconducting atom chip $C$ of an elongated~\cite{MX_ARIMONDORYDLATT11} $^{87}$Rb atom thermal cloud cooled below 1~$\mu$K~\cite{ENS_SUPERCONCHIP06,ENS_BECCHIP08}, near quantum degeneracy. This sample is trapped in a red-detuned focused laser beam and brought inside the `science' capacitor $S$. We suppress the ground state trap and laser-excite a low angular momentum $n=50$  Rydberg state in the dipole blockade regime, leading to a random Rydberg atom chain ($\simeq 110$ atoms) with inter-atomic spacings of the order of 9~$\mu$m~\cite{ENS_CHIPINTERACTION15,ENS_CHIPSPECTRO14}. We get rid of the residual ground-state atoms with a resonant pushing laser pulse and transfer the Rydberg atoms into $\ket{50C}$  using a $\sigma^+$-polarized evanescent rf field. During the few microseconds required for this sequence, atomic motion is negligible.

The Laguerre-Gauss radial confinement beam is then switched on. We also switch on two 1~$\mu$m-wavelength `plug' Gaussian beams parallel to $OY$. They create two energy barriers on the $OX$ axis, centered at $X=\pm L/2$. The `right' plug ($X=L/2$) is lower than the `left' one. We then slowly compress the trap by reducing $L$. We increase accordingly the van der Waals repulsive interaction up to a point where the energy of the right-end atom compares to that of the weak plug. Further compression ejects atoms, one at a time, above the weak plug. The `evaporation' of an atom removes a part of the global energy, providing a cooling mechanism reminiscent of the evaporative cooling~\cite{MX_KLEPPNEREVAP88}. The final atom number, $N$, is determined by the height of the weak plug and by the final value of $L$. 

\begin{figure}
\begin{center}
\includegraphics[width=8cm]{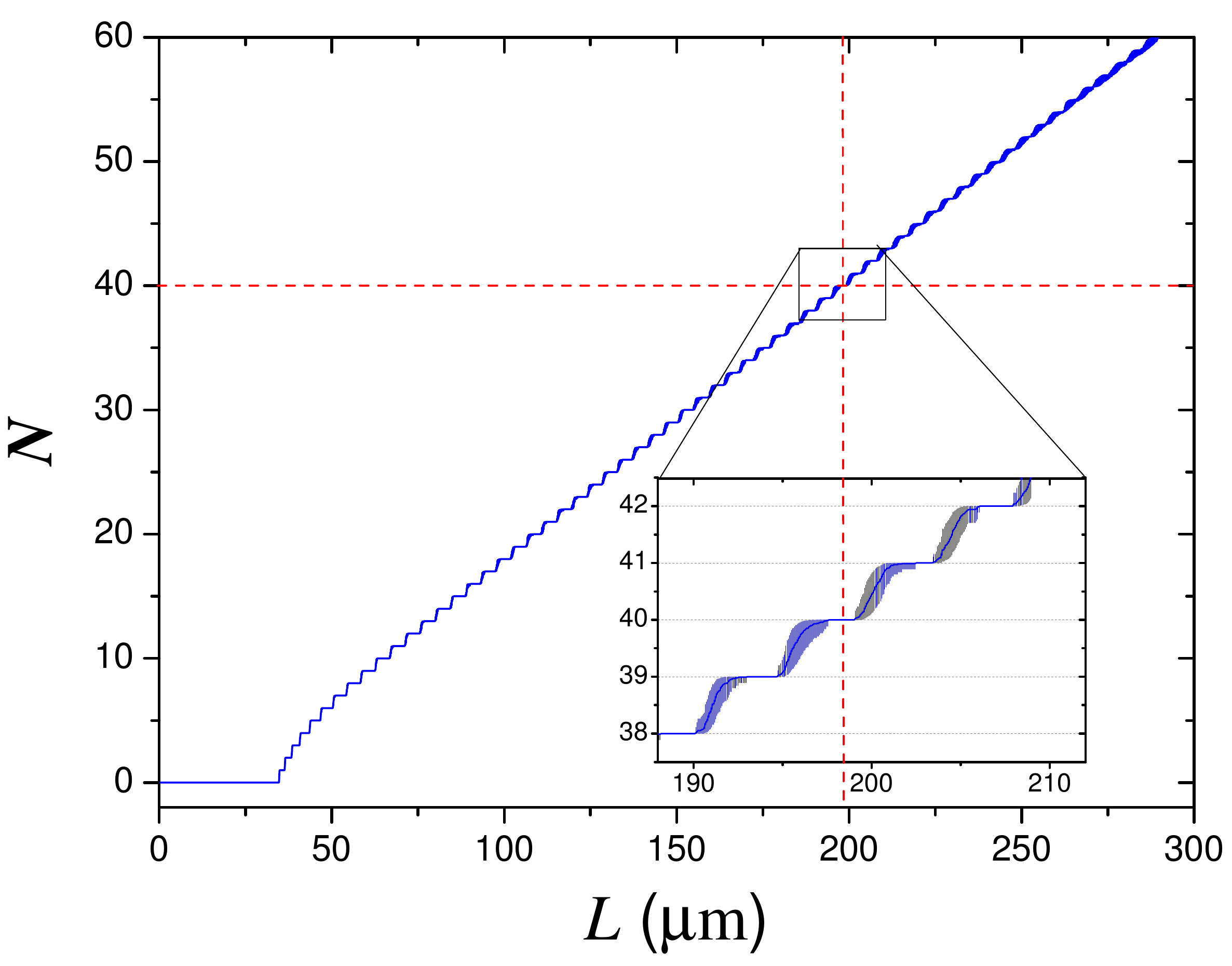}
\end{center} 
\caption{Number of atoms left as a function of the distance $L$ between the two plug beams. The thick curve gives the average over 100 realizations of the evaporation process. The atom number variance is indicated by the blue-shaded area.}
\label{fig:nevap}
\end{figure}

Numerical simulations of the classical atomic dynamics reveal the efficiency of this process. Figure~\ref{fig:nevap} presents the average and the variance over 100 realizations of the evaporation sequence of the number of remaining atoms as a function of the final $L$ value. For atom numbers lower than 45, we observe clear steps in the evolution of $N$. The zoom around $N=40$ (inset) shows that the atom number variance cancels for optimal $L$ values. Stopping the evaporation process at such trap lengths deterministically prepares a string with a prescribed atom number. The interatomic spacing is finely tuned through a final adjustment of $L$. The lattice is then adiabatically turned on, trapping the atoms in their respective sites (the plugs remain on with an adjusted power to compensate the repulsion of the end atoms by their single neighbor).

The complete preparation sequence simulated here lasts  1.3~s (Appendix~\ref{app:evap}). In order to avoid atomic decay during this relatively long time interval, the electric field $F$ can be raised to a large value, leading to an individual atom lifetime $> 200$~s. The final longitudinal position dispersion with respect to the lattice sites is $\Delta X=65$~nm for $N$=14 atoms, corresponding to only $\approx 1$ oscillation quantum (110~nm for $N=40$, i.e. $\approx 4$ quanta). A full quantum model would be clearly required. It is out of the scope of this paper, and   $\Delta X$ will be used in the next Section for an order of magnitude estimate of the influence of the atomic motion. We have checked with 3-D simulations of the dynamics that the transverse position dispersions $\Delta Y=\Delta Z$ are of the same order of magnitude as $\Delta X$.

The evaporation procedure can also be used for an efficient detection of the spin states. At the end of the spin-chain evolution, the exchange interaction can be halted by casting with a `hard' microwave $\pi$-pulse $\ket{48C}$ onto $\ket{46C}$. The exchange interaction $\ket{46C}\leftrightarrow\ket{50C}$ is in the mHz range. The energy states of the spins are thus frozen from then on. The repulsive  van der Waals interactions being nearly unchanged, the evaporation process can be resumed. The lattice is switched off, the right plug is lowered, and $L$ is slowly decreased, expelling atoms one at a time. The atoms escape along the $OX$ axis, guided by the LG beam at a velocity determined by the height of the weak plug. They fly towards the field-ionization region ($D$ on Fig.~\ref{fig:compscheme}). The levels $\ket{50C}$ and $\ket{46C}$ are selectively detected there with near-unit detection efficiency. This simple scheme reads out the spin states in the up/down basis. Adding a hard microwave pulse before freezing the interaction, we can rotate the equivalent spin at will and thus detect any spin observable (the same for all atoms) and its correlation functions along the chain. Microwave pulses acting on individual atoms on their way from $S$ to $D$ make it possible in principle to measure arbitrary quantum observables of the spin chain.

The ability to measure, as a function of time, the states of the individual spins opens a wealth of possibilities. It is instrumental to access complex correlation functions and entanglement properties in the spin chain.

\section{Numerical simulation of adiabatic evolution through a quantum phase transition}
\label{sec:simus}

In this Section, we discuss the observation of quantum phase transitions using this setup. In particular, we investigate the influence of the residual atomic motion around the lattice sites. We include the effect of the classical atomic motion in the spin-chain Hamiltonian discussed in Section~\ref{sec:hamiltonian}  and in the numerical simulations of the system dynamics. This effect is quite dependent upon the relative values of the exchange frequency $J$ and of the trap oscillation frequency $\omega_X$. We thus explore numerically the two cases, $d=5\ \mu$m and $d=7\ \mu$m, corresponding respectively to $J\approx\omega_X$ and to $J\ll\omega_X$.

\subsection{Hamiltonian with a classical atomic motion}

We treat the atomic motion as classical and independent from the spin dynamics. We use the results of the numerical simulations of the evaporation process (Section~\ref{sec:chain} and Appendix~\ref{app:evap}) as an input for the atomic trajectories and perform averages over the outcomes of many (100) realizations of the evaporative chain preparation. The Hamiltonian including the atomic motion can be written as 
\begin{widetext}
\begin{equation}
\frac{H}{h} = \sum_{j=1}^N\left[ \frac{\nu_0-2\nu}{2} \sigma^z_j  + \frac{\Omega}{2} \sigma^x_j \right]
+\sum_{j=1}^{N-1} I_{j,j+1}(t)\left[J_{z}\,\sigma^z_j\sigma^z_{j+1} + J\left(\sigma^x_j\sigma^x_{j+1} + \sigma^y_j\sigma^y_{j+1}\right) 
+ \frac{\delta\zeta}{2}\left(\sigma^z_j + \sigma^z_{j+1} \right)
 \right]\;;
\label{eq:realisticHam}
\end{equation}
\end{widetext}
We have introduced
\begin{equation}
I_{j,j+1}(t) = \frac{d^6}{[x_{j+1}(t) - x_j(t)]^6},
\label{eq:couplings}
\end{equation}
where $x_j$ is the position of atom $j$. A regular lattice ($x_j=jd$ within a constant offset) corresponds to $I_{j,j+1} = 1$.
An important remark is that, even though the absolute strengths of the coupling coefficients  fluctuate with position and time, the ratios of these couplings  are constant, in time and along the chain. The motion induces thus a highly correlated noise on the couplings.

\begin{figure*}
\centering
\includegraphics[width=2\columnwidth,clip]{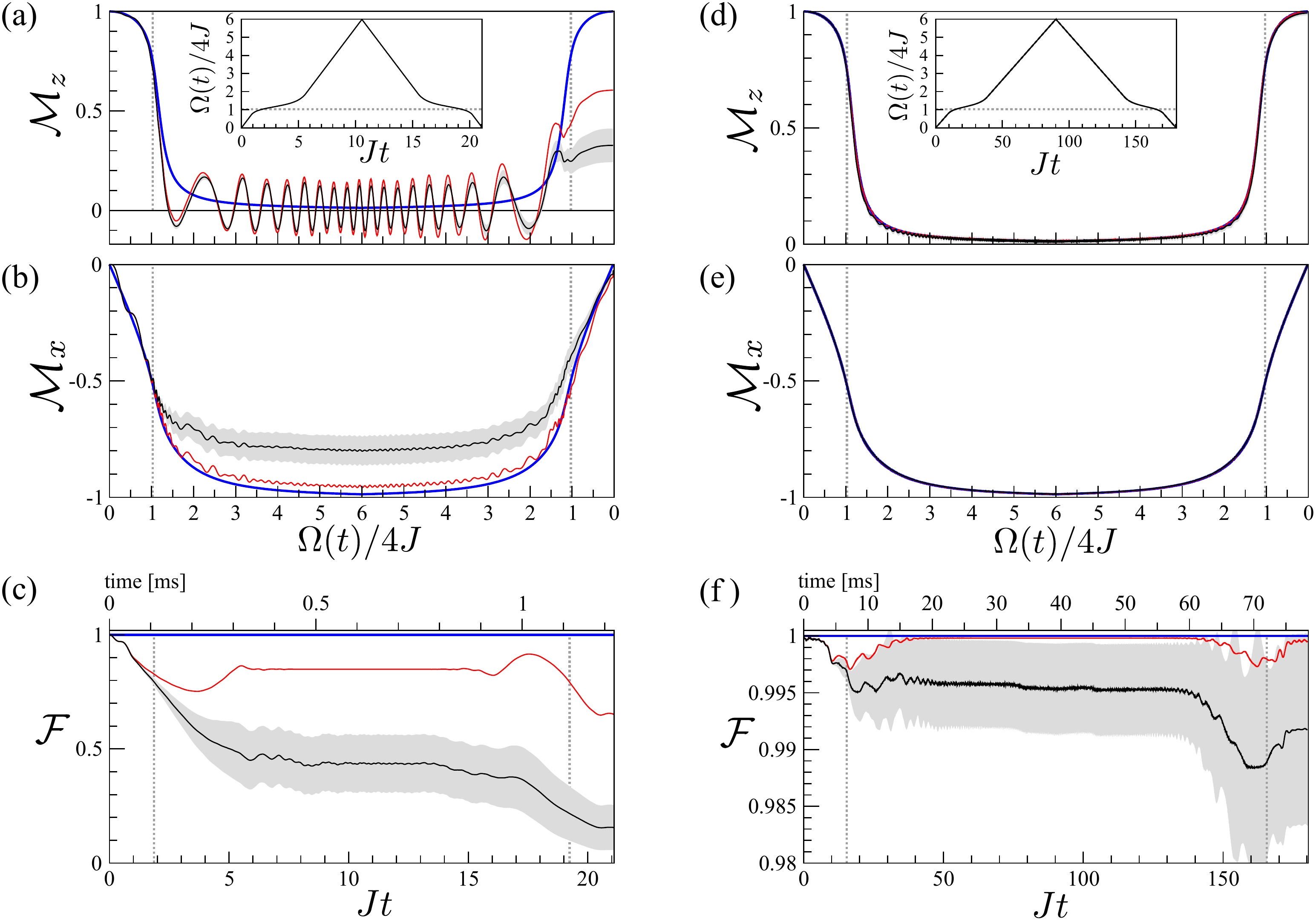}
\caption{\textbf{Simulation of an adiabatic preparation of the ground state for the ferro-para transition at $J_z/J= -1.6$ and with $N=14$ spins.} (a), (b) and (c) total magnetizations $\mathcal{M}_z$, $\mathcal{M}_x$ and fidelity $\mathcal{F}(t)$ for  $J=17$~kHz. (d), (e) and (f) same curves for  $J=2.3$~kHz [note that the  time scale and vertical scale for the fidelity differ from those of frames (a-c)]  The insets in frames (a) and (d) depict the optimized ramp $\Omega(t)$. For each quantity $O$, the black curve gives its average over 100 realizations of the classical atomic trajectories.  The shaded area represents the corresponding standard deviation of the distribution over trajectories. The red curves corresponds to a situation with atoms fixed at the lattice sites. The blue curves correspond to the exact ground state of the Hamiltonian for motionless atoms.  In each frame, the vertical/horizontal dotted lines correspond to the expected quantum phase transition point.}
\label{fig:quench}
\end{figure*}
	
The $\delta\zeta$ term adds a random longitudinal magnetic field along the $z$-direction. We chose the dressing frequency $\nu$ so as to cancel the average value of this field:  $\overline\Delta=\nu_0- 2\nu +2\delta \zeta \overline{I}=0$, where $\overline{I} \simeq 1$ and the over-line denotes the  average over many realizations of the atomic trajectories.
Still, a residual magnetic field, $\overline{\Delta'}=\nu_0- 2\nu + \delta\zeta\overline{I} = -\delta\zeta_0\overline{I}$, remains on the two edge sites $j=1,N$. This field breaks locally the $\mathbb{Z}_2^z$ symmetry and polarizes the edge spins in the $z$-direction. It is an asset or a drawback depending on the purpose of the quantum simulator. It is, for instance, an asset while entering a ferromagnetic phase. It creates a perturbation that naturally triggers the build-up of the order parameter. Note that, for large enough chains and in gapped phases, these edge effects are relevant only over the correlation length scale. The physics of the model can still be captured anyway in the bulk of the chain.

\subsection{Adiabatic evolution through a quantum phase transition}

We now investigate the evolution of the system in an adiabatic evolution through a quantum phase transition line. We perform simulations of the full system dynamics under the Hamiltonian~(\ref{eq:realisticHam}) for up to $N=14$ atoms using exact diagonalization. We infer, from Fig.~\ref{fig:numphasediag}, that a favorable situation to probe a quantum phase transition is the ferro-para transition F$\leftrightarrow$P$_x$. It has little finite-size effects and a strong ordering in the ferromagnetic phase. We take thus $J_{z}/J = -1.6$, which corresponds to $F=6$V/cm and $B = 14$ Gauss in Fig.~\ref{fig:jzj} and leads to $\delta\zeta/J \simeq 1.68$.

The edge fields are then negative and favor the spin-up ferromagnetic state $\ket{\text{F}}=\otimes_{j=1}^N\ket{\uparrow}$. This state is actually the ground state of \eqref{eq:realisticHam} for $J_{z}/J = -1.6$ and $\Omega=0$ and can be straightforwardly prepared experimentally. We note that, in the opposite limit,  $\Omega\gg 4J$, the polarized state $\ket{\text{P}_x}=\otimes_{j=1}^N\ket{\leftarrow}$, where $\ket{\leftarrow}=(\ket{\uparrow}-\ket{\downarrow})/\sqrt{2}$ become the ground state.
Starting from an exact ground state is an ideal situation for an adiabatic preparation protocol.
We choose thus to start from $\ket{\text{F}}$, to vary $\Omega(t)$ from $0$ to $\Omega_\text{max}/4J=6$ and then to decrease $\Omega$ by reversing the $\Omega(t)$ function.
This protocol has two goals. First, we follow the behavior of the observables along the path in order to probe the transition and, second, this cycle allows us to probe the deviations from adiabaticity through the comparison between the observables in the direct and return ways.

Adiabatic theory suggests~\cite{Roland2002,Kim2011} to use non-linear ramps for $\Omega$, with a velocity $\dot\Omega$ proportional to the square of the gap to the first excited states. In the presence of motion, we phenomenologically found good non-linear ramps with a velocity inversely proportional to the derivative $d\mathcal{M}_z/d\Omega$ calculated in the ground state (very low velocity values are replaced by a constant lower bound). In order to save computing time, we first optimize the ramps using simulations for  $N=10$ spins and reuse them for the largest calculation ($N=14$).

We plot on Fig.~\ref{fig:quench}, for a  $N=14$ spins chain,  the average values and the standard deviation (over 100 atomic motion realizations) of $\mathcal{M}_z$ and $\mathcal{M}_x$. We also plot the fidelity of the  time-evolving state, $\ket{\psi(t)}$, with respect to the ideal ground state $\ket{\psi_0(\Omega)}$ for a given $\Omega$: $\mathcal{F}(t) = |\braket{\psi(t)}{\psi_0(\Omega(t))}|^2$. Frames (a-c) correspond to $J=17$~kHz, frames (d-f) to $J=2.3$~kHz. The optimized ramps $\Omega(t)$ are given in the insets of frames (a) and (d). The realistic averaged curves (black lines) are compared to the ground state (blue lines) and to the time evolutions obtained with the same time-dependent protocol operating on atoms at fixed positions (red lines).

In the thermodynamic limit, the transition (indicated by the vertical dashed lines in Fig.~\ref{fig:quench}) would be signaled by a vanishing of $\mathcal{M}_z$ at the critical point and a discontinuity in the slope of $\mathcal{M}_x$, both with critical exponents belonging the Ising universality class. On a finite chain, the transitions are smoothed out.
The data of frames (a-c) in Fig.~\ref{fig:quench} clearly exhibit, for $J=17$~kHz, the expected behavior of the magnetization observables around the phase transition points. However, imperfections are conspicuously revealed by the intermediate oscillations in $\mathcal{M}_x$ and the reduced final value of $\mathcal{M}_z$ (which, in principle, should return to its initial value, 1). The protocol generates ``heating'', mostly close to the transition points, and the fidelity $\mathcal{F}$ accordingly sharply drops at the transition.

 Part of these imperfections are due to the atomic motion, as shown by the differences between the black and red curves. These motion-induced imperfections increase rapidly when the sweep time is increased. We are thus driven to use a rather fast ramp (the total duration $T=1.2$~ms of the sequence corresponds to $JT=20$ only). Accordingly, part of the imperfections are due to the breaking of the adiabaticity criterion, as illustrated by the difference between the red and blue curves.

A lower $J$ value (2.3~kHz) leads to a considerably improved situation, as shown in frames (d-f) of Fig.~\ref{fig:quench}. The atomic motion is effectively decoupled  from the spin dynamics. This decoupling allows us to use a much slower ramp. The total duration $T=79$~ms corresponds now to $JT=180$. The differences of the observables in the three situations are then negligible. The final fidelity of the 14-spin state reaches an outstanding value of 0.99.

These preliminary results show that it is fairly easy to achieve operating conditions such that the residual classical atomic motion has a quite negligible influence on the spin dynamics. The long lifetime of the spin chain is instrumental to realize slow evolutions fulfilling the adiabaticity criterion.  This would allow us to explore properly the complete phase diagram and the quantum phase transition phenomenon.  Obviously, further studies could lead to further optimizations of the ramps making it possible to operate at larger couplings over a reduced time scale and to the exploration of the other transitions in the phase diagram.

\section{Conclusion}
\label{sec:conclusion}

We have shown that state-of-the-art techniques make it possible to build a spin-chain quantum simulator based on laser-trapped circular Rydberg atoms. This simulator combines the flexibility of atomic lattices, the individual atomic observables read-out typical of ion trap together with the strong dipole-dipole interactions of Rydberg atoms. Defect-free atomic chains can be prepared by an evaporative cooling method, which leaves the atoms finally near their vibrational ground state. Evaporation also provides us with a unit efficiency individual spin detection. A proper microwave dressing leads to a fully tunable spin-$1/2$ XXZ chain Hamiltonian. Its parameters are under direct experimental control, a unique feature of this simulator. The long lifetime of the laser-trapped circular atoms, protected from spontaneous emission, makes it possible to follow the dynamics over unprecedented time intervals, in the range of  $10^5$ times the spin flip-flop period. Moreover, the individual detection of all spin observables makes it possible to access a wealth of interesting properties, such as entanglement properties and local entropies.

\remodified{Let us stress that the techniques proposed in this paper could have a deep impact on the thriving Rydberg atoms physics, well beyond the realization of a full-fledged quantum simulator. Many experiments are considerably hindered by the lack of trapping of the Rydberg atoms and by the spurious transfers induced by blackbody radiation. We propose here simple solutions to overcome these bottlenecks. For instance, Rydberg atoms have been shown to be ultrasensitive probes of their electromagnetic environement~\cite{MX_PFAUEFIELDRYD12,MX_SHAFFERELECTROMETRY13,MX_DUMKERYDBERGSURF14,MX_RAITHELMICROWAVEEIT14,ENS_RYDELEC16}. Adding to these experiments the laser trapping capability, compatible with all high-angular-momentum states, would allow the realization of extremely sensitive, well-localized probes of the local fields. Cavity quantum electrodynamics could also considerably benefit from the techniques outlined here. Rydberg atoms cavity QED experiments have been plagued by the lack of deterministic atom sources and by the fast transit of the thermal atoms across the cavities~\cite{ENS_RMP}. Laser trapping allows to remove straightforwardly these bottlenecks. One can even envision hybrid cavity QED experiments combining superconducting circuits and laser-trapped Rydberg atoms, which can be used to create a coherent interface between microwave and optical photons~\cite{ENS_CIRCULARIZATION17}.}

\remodified{Returning to quantum simulation, }a circular state simulator with about 40 atoms could address important problems of many-body quantum physics. We have shown that slow variations of the Hamiltonian parameters make it possible to explore precisely the quantum phases of the XXZ model, generating the ground states with a high fidelity. \modified{Of course, these ground states are well-known and most of their properties can be assessed using standard numerical techniques, as the DMRG used in our extensive numerical computations. Checking the agreement between the observed phases with the expectations mostly assesses the quality of the simulator. It would in particular show that the residual atomic motion and other experimental imperfections have a negligible influence, as we expect.}

\modified{The real interest of this simulator lies in studies of the spin chain dynamics, much more demanding numerically when highly excited states or slow evolutions are at stakes~\cite{MX_BLOCHMBL17}. For instance, 
fulfilling the adiabatic limit in a transition towards a gapless phase is more and more difficult when the systems size increases. A too fast crossing of  the transition line results in the generation of defects with respect to the theoretical final ground state. Exploring the generation of these defects and the limits of the adiabatic regime  is particularly important in the context of adiabatic quantum computation~\cite{Chandra2010}, quantum annealing~\cite{QI_DWAVEANNEAL14,QI_TROYERANNEAL15} and Kibble-Zurek mechanism~\cite{MX_KIBBLEZUREK05}.}

An essential perspective for such ground-state physics is to explore the spin-one Haldane phase~\cite{Haldane1983,Haldane1983a} using the ladder geometry. Separately prepared parallel chains could be brought in interaction (by moving their Laguerre-Gauss transverse trapping beams), leading to a square ladder geometry. Using the anisotropy of the dipole-dipole interaction, the signs of the coupling between legs (along $OX$) and rungs (along $OZ$) of a properly oriented ladder can be different. This leads to two antiferromagnetic chains that are ferromagnetically coupled. This model, in part of its phase diagram, realizes the Haldane phase~\cite{Narushima1995,MX_WHITECHAIN96,Hijii2005}. This phase possesses a non-trivial topological order~\cite{Nijs1989}, which can be straightforwardly measured in this context, and fractional spin-1/2~edge states with the open boundary conditions typical of our simulator~\cite{Hagiwara1990}. \modified{ There also, the ground states and low excitations physics of the system are well apprehended by numerical methods, but the dynamical evolution is much less easy to simulate.}

Another interesting low-energy physics problem is that of a disordered XXZ chain~\cite{Ma1979,Dasgupta1980,Doty1992,Fisher1994,Igloi2005}.
Adding a laser speckle field to the optical lattice, it is fairly easy to produce random shifts of the atoms with respect to their equilibrium positions, randomly modulating the dipole-dipole interactions.
In the $J_z<0$ regime, this model displays the paradigmatic competition between localization and interactions, \modified{ a subject of an intense activity in quantum simulation} opening the way for Bose-glass physics~\cite{Giamarchi1987,Giamarchi1988,Giamarchi2004}.
Another striking feature of this model is the emergence of random singlet phases~\cite{Ma1979,Dasgupta1980,Doty1992,Fisher1994,Igloi2005}, with their unusual long range correlations and entanglement properties~\cite{Refael2004,Laflorencie2005} in disordered systems. 
Remarkably, the random singlet phase of the Heisenberg point would be accessible thanks to the possibility to tune $J_z=J$ on all bonds. 

The ability to modulate rapidly the Hamiltonian parameters also opens a vast realm of possibilities~\cite{Silveri2017}. Periodic modulations could be used to realize spectroscopic investigations of the elementary excitations of the system. They bear a particular interest at the critical point of the Ising transition (the one studied in Section~\ref{sec:simus}), as shown by its remarkable integrable features~\cite{Zamolodchikov1989,Kjaell2011}, recently investigated in condensed matter experiments~\cite{Coldea2010,Morris2014}. The long lifetime of the circular simulator would be instrumental in studying low-frequency excitations, not easily accessed in other contexts. 

Floquet engineering corresponds to periodic variations of the couplings much faster than $J$. It allows one to design effective Hamiltonians that are not accessible with the usual control parameters~\cite{MX_ARIMONDOLATTSHAKE07,MX_KOLOWSKYFLOQUET11,MX_DALIBARDGAUGE14}. This is a particularly interesting perspective to enlarge the field of applications of the circular state quantum simulator, since all parameters of $H$ can be easily modulated at high frequencies. In the same spirit, the proposed Rydberg set-up notably makes it possible to study Floquet time crystals~\cite{MX_ELSETIMECRY16,Sacha2017}.

Instantaneous quenches can be realized by a sudden variation of the Hamiltonian. There is a whole range of questions on quenches that would benefit from long observation times. Whether an isolated quantum system displays equilibration and thermalization is a fundamental issue of statistical physics~\cite{Dziarmag2010,Polkovnikov2011,DAlessio2016,MX_BORGNOVITHERM16,QC_MARTINISIMUL16}. As the spin-chain Hamiltonian has integrable points, one could investigate the interplay between thermalization and integrability~\cite{Kinoshita2006}. The intermediate relaxation time regime contains information on the propagation of correlations at the origin of the relaxation process~\cite{MX_BLOCHCORR12}. Another remarkable scenario is the pre-thermalization~\cite{MX_SCHMIEDMEYERTHERMAL16}. Some observables reach rapidly a metastable steady-state, while the system is not yet in its thermal equilibrium. Only few experiments have been carried out in this regime~\cite{Gring2012}. Finally, the dephasing time of a sub-system could be directly measured~\cite{MX_BARTHELDEPHASE08}.
Combining quench protocols with disordered Hamiltonians offer a way to address the issues related to many-body localization~\cite{MX_BASKO06,MX_HUSEMBL96,Nandkishore2015,Lueschen2016}. In particular, the long simulation times would allow one to follow the logarithmic increase of the entropy that signals the many-body localization transition~\cite{MX_ZNIDARICMBL08}.

Beyond the spin chain physics, the circular state simulator could explore a new regime of spin-boson interaction~\cite{Leggett1987,LeHur2008,Porras2008}. Shallow optical lattices lead to a situation in which the spin exchange is strongly coupled to the atomic motion~\cite{MX_KIMBLESPINCHAIN17}. The joint motion of the atoms would then entangle with the spins, leading to a situation, in which numerical simulations are far out of reach even for moderate spin numbers. In particular, the common coupling of the spin ensemble to the same bath could mimic correlated errors, which are one of the key problems for quantum error correction in quantum information protocols.

\modified{ We have limited our discussions to chains with even couplings, since this is the first interesting problem that this simulator could address. The use of recent atom trapping techniques could considerably extend the simulation realm. Transposing to this context the individually controlled multiple atomic traps on  a line demonstrated in~\cite{ArXiv_Bernien2017}, would allow us to individually control the position of each atom. We may thus envision preparing, from an appropriate lattice, a linear chain of circular atoms with rather large separations (10 $\mu$m or more) making interactions negligible. We could then move (in a time of the order of the trap oscillation frequency, using optimal control techniques) the atoms to put them, by pairs, in interaction for a given set of time, during which the dressing source and static field can be adjusted to provide individually controlled pair interaction parameters. \remodified{We could use this technique, for instance, to get rid of the spurious next-nearest-neighbor coupling if it has undesired effects on the simulation.  We can thus envision a simulator of a }completely general spin-spin interaction model, with complete single-site addressing, both at the interaction and at the detection stage.}

\modified{Finally, extensions to full 2-D or even 3-D geometries can also be envisioned, also by a mere extension to this context of recent techniques for atomic lattices and programmable optical tweezers~\cite{MX_BROWAEYS2DRYD16}. \remodified{One could, for instance, prepare a few defect-free chains of Rydberg atoms by the van der Waals evaporation method, and then pick out each atom  with individual laser tweezers to bring them finally in an arbitrary spatial arrangement. The extremely long atomic lifetimes and the tight laser trapping of the circular atoms makes such a scheme feasible. The techniques involved  are demanding, but already well established in other contexts}. This dramatic extension of the proposed quantum simulator capability would allow it to address a domain where understanding the mere ground state is already quite challenging, not to mention the long-time-scale dynamics. We thus think there is a bright long-term future for circular-state quantum simulators.}

\acknowledgements
We acknowledge funding by the EU under the FET project `RYSQ' (ID: 640378) and by the ANR under the project `TRYAQS' (ANR-16-CE30-0026). We are grateful to B. Dou{\c c}ot, Th. Giamarchi, Ph. Lecheminant, D. Papoular and P. Zoller for fruitful discussions.

\appendix
\section{Circular states and their van der Waals interaction}
\label{app:vdW}

\begin{figure}
\begin{center}
\includegraphics[width=7cm]{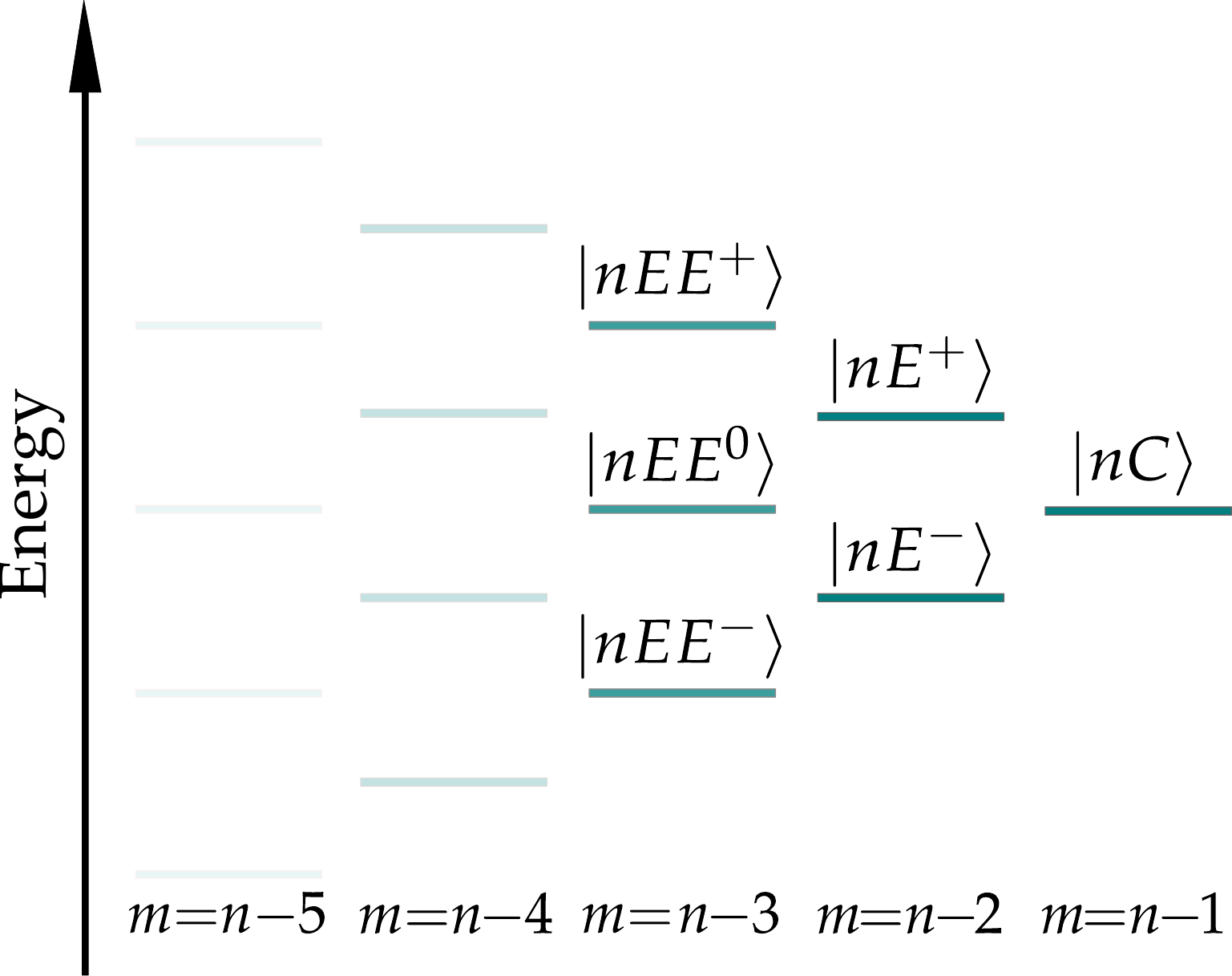}
\end{center} 
\caption{Diagram of the Stark levels with the highest magnetic quantum numbers. The circular state $\ket{nC}$ is at the tip of the triangle of Stark levels sorted according to $m$.}
\label{fig:levels}
\end{figure}

For Rydberg levels with a high angular momentum, the quantum defects are negligible and the hydrogenic model is an excellent approximation. In vanishing electric and magnetic fields, the circular state with principal quantum number $n$, $\ket{nC}$, is degenerate with the enormous hydrogenic manifold.  Any perturbation admixes it with other high-$\ell$ `elliptical' states~\cite{ENS_DAHU}. In a static electric field, the  manifold degeneracy is partially lifted~\cite{TXT_GALLAGHER}. The eigenstates of the Stark Hamiltonian in an electric field $\mathbf{F}$ along $OZ$ can be sorted out by their magnetic quantum number $m$ ($\ell$ is no longer a good quantum number since the spherical symmetry of the Hydrogen atom is broken). The energy spectrum of the manifold arranges as a triangle whose tip is the circular level $\ket{nC}$, as shown in Fig.~\ref{fig:levels}, isolated from the nearest elliptical states  $\ket{nE^\pm}$. A magnetic field $\mathbf{B}$, also along $OZ$, lifts the near-degeneracy of $\ket{nC}$ with $\ket{nEE^0}$. The circular state is now stable against stray field perturbations. The circular level experiences a negative second-order Stark shift, scaling as $n^6$, $-1.582$~MHz/(V/cm)$^2$ for $n=48$. The differential Stark shift on a transition between two circular states is much lower [$-438$~kHz/(V/cm)$^2$ on the two-photon $ \ket{48C}\rightarrow\ket{50C}$ transition]. 

Due to their high angular momentum, circular states cannot be reached directly by laser excitation of the ground state. Their preparation relies on the laser excitation of a low-$\ell$ Rydberg state, followed by a series of $\sigma^+$-polarized radio-frequency transitions between Stark levels, performed in an adiabatic rapid passage sequence~\cite{ENS_CIRCRB}. A good control of the radio-frequency field polarization leads to an efficient ($\simeq 95$~\% efficiency and purity) and rapid (few $\mu$s) transfer into the circular state~\cite{ENS_QZDEXP14}. Field-ionization provides a state-selective detection with near unit efficiency~\cite{ENS_COUNT05}.

For a pair of interacting Rydberg atoms at a distance $d$ along $OX$, perpendicular to the quantization axis $OZ$, the dipole-dipole interaction reads
\begin{eqnarray}
V_{dd} = \frac{e^2 r_1 r_2}{3\epsilon_0 d^3} &\Big[&Y_1^{0}Y_1^{0} +\frac{1}{2}\left(Y_1^{+1}Y_1^{-1}+Y_1^{-1}Y_1^{+1}\right)-\nonumber\\&&\frac{3}{2}\left(Y_1^{+1}Y_1^{+1}+Y_1^{-1}Y_1^{-1}\right) \Big] \ ,
\label{eq:vdwinter}
\end{eqnarray}
where $r_1$ and $r_2$ are the distances of the two Rydberg electrons to their respective cores and where the $Y_i^j$ are the spherical harmonics for the two electron positions.  

We encode the spin-up and spin-down states of the simulator on the $\ket{50C}$ and $\ket{48C}$ circular states, connected by a two-photon transition at frequency $\nu_0=111.95$~GHz. In the basis $\{\ket{48C,48C}$, $\ket{48C,50C}$, $\ket{50C,48C}$, $\ket{50C,50C}\}$,  the dipole-dipole interaction reads, in a second order perturbative approximation,
\begin{equation}
 V= \frac{h}{d^6} 
 \begin{pmatrix}
 C_{6,48-48} & 0 & 0 & 0 \\
0 & C_{6,48-50} & A_{6,48-50} & 0 \\
0 & A_{6,48-50} & C_{6,48-50} & 0 \\		
0 & 0 &0 & C_{6,50-50}		
\end{pmatrix}\ .
\end{equation}

In terms of the Pauli operators for the two atoms, $\sigma^{x,y,z}_i\  (i=1,2)$, this interaction can be rewritten as
\begin{equation}
\frac{ V}{h} =\delta E\openone+\frac{\delta\zeta}{2}\left(\sigma^z_1 + \sigma^z_2\right) +J_z\,\sigma^z_1\sigma^z_2 +J\left(\sigma^x_1\sigma^x_2+\sigma^y_1\sigma^y_2\right) \ ,
\end{equation}
where
\begin{eqnarray}
\delta E&=&\frac{C_{6,48-48}+C_{6,50-50}+2A_{6,48-50}}{4d^6}\ ,\\
\delta\zeta &=&\frac{C_{6,48-48}-C_{6,50-50}}{2d^6}\ , \\
J_z &=&  \frac{C_{6,48-48}-2C_{6,48-50}+C_{6,50-50}}{4d^6}\ ,\\ 
J &=&\frac{|A_{6,48-50}|}{2d^6} \ .
\end{eqnarray}
Note that the sign of the exchange term $J$ is irrelevant since it can be changed by a mere redefinition of the absolute phase of the basis levels. We thus chose it to be positive. The $\delta E $ term is a mere redefinition of the energy origin, that will no longer be explicitly included in our discussions. The $\delta\zeta$ term results from the differential van der Waals shift between the two atomic levels and plays the role of a longitudinal  field in the spin model. The $J_z$ and $J$ terms describe the longitudinal and transverse (exchange) spin-spin interactions respectively.

In order to determine precisely these coefficients, we perform an explicit numerical diagonalization of the pair Hamiltonian, including the Zeeman and Stark perturbations (note that the dipole-dipole interaction breaks the cylindrical symmetry of the Stark levels in the proposed geometry, preventing us from using approximate analytical solutions). We have to restrict the total Hilbert space in order to perform the computation. We limit  its basis to levels whose principal quantum numbers differ by $|\Delta n|<3$ from 48 or 50 (the coupling matrix elements decrease rapidly when $\Delta n$ increases). We also select $m$ values differing by at most $|\Delta m|<3$ from those of the levels or interest. Most of the computations are performed with a basis of 361 pair states. For a few values of the fields, we have checked that the interaction changes by only $\simeq 1$\% when using a three times larger basis.

We have first computed the interaction between two atoms in $\ket{50C}$ as a function of the interatomic distance, for $B=13$~Gauss and $F=6$~V/cm. The uncoupled $\ket{50C,50C}$ pair state is found to be mainly contaminated by the $(\ket{50E^+,50E^-}+\ket{50E^-,50E^+})/\sqrt 2$ symmetric pair state.  The energy variation of the levels is in excellent agreement with a $1/d^6$ dependence for $d>3$~$\mu$m. For smaller distances, the interaction is too large to agree with the perturbative van der Waals dependence.

For  $d=5\ \mu$m, we find  $A_{6,48-50}=-0.539$~GHz\,$\mu$m$^6$, a value independent (within $10^{-5}$) of the electric and magnetic fields in the relevant range. The other $C_6$  coefficients have a marked dependency on $F$ and $B$, varying by 10 to 20\% for $6<F<12$~V/cm and $9<B<16$~Gauss. Their values for $F=9$~V/cm and $B=13$~Gauss are $C_{6,48-48}=2.2$~GHz\,$\mu$m$^6$,  $C_{6,48-50}=2.66$~GHz\,$\mu$m$^6$ and $C_{6,50-50}=3.03$~GHz\,$\mu$m$^6$.

\begin{figure}
\begin{center}
\includegraphics[width=7cm]{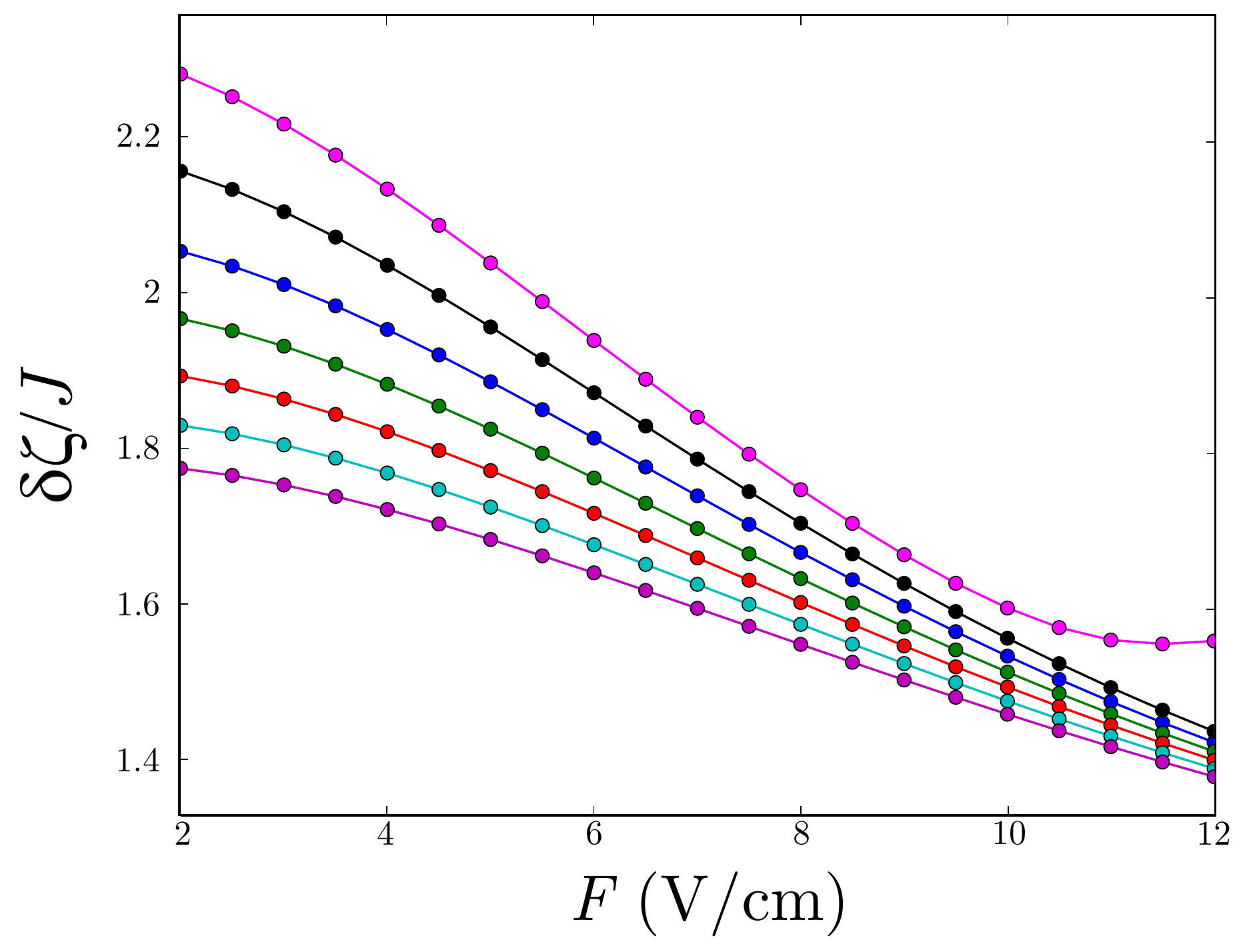}
\end{center} 
\caption{Variations of  $\delta\zeta/J $ as a function of the electric field $F$ for $B=9,10,11,12,13, 14$ and $15$~Gauss (magenta, black, blue, green, red, cyan and purple dots respectively). The colored lines are a guide to the eye.}
\label{fig:jzvar}
\end{figure}

Accordingly, in terms of the spin model, $J=17$~kHz at $d=5\ \mu$m (2.3~kHz at 7~$\mu$m) is independent of the fields, whereas $J_z$ and $\delta\zeta $ vary over large ranges. Figure~\ref{fig:jzj} shows the variations of $J_z/J$  as a function of $F$ and $B$. Fig.~\ref{fig:jzvar} shows the corresponding variations of  $\delta\zeta/J $. Note that $J_z/J$ and $\delta\zeta/J $ do not depend upon $d$. We observe that the $J_z$ dependence flattens when $B$ increases. On the other hand, a larger $B$ value reduces the mixing of the circular states and the elliptical states, and accordingly increases the levels lifetime (Appendix~\ref{app:losses}). We thus chose the largest $B$ value for which the spin chain can be tuned over the complete phase diagram, $B=13$~Gauss.

\section{Details on numerical simulations}
\label{app:numerics}

\begin{figure}[t]
\centering
\includegraphics[width=\columnwidth,clip]{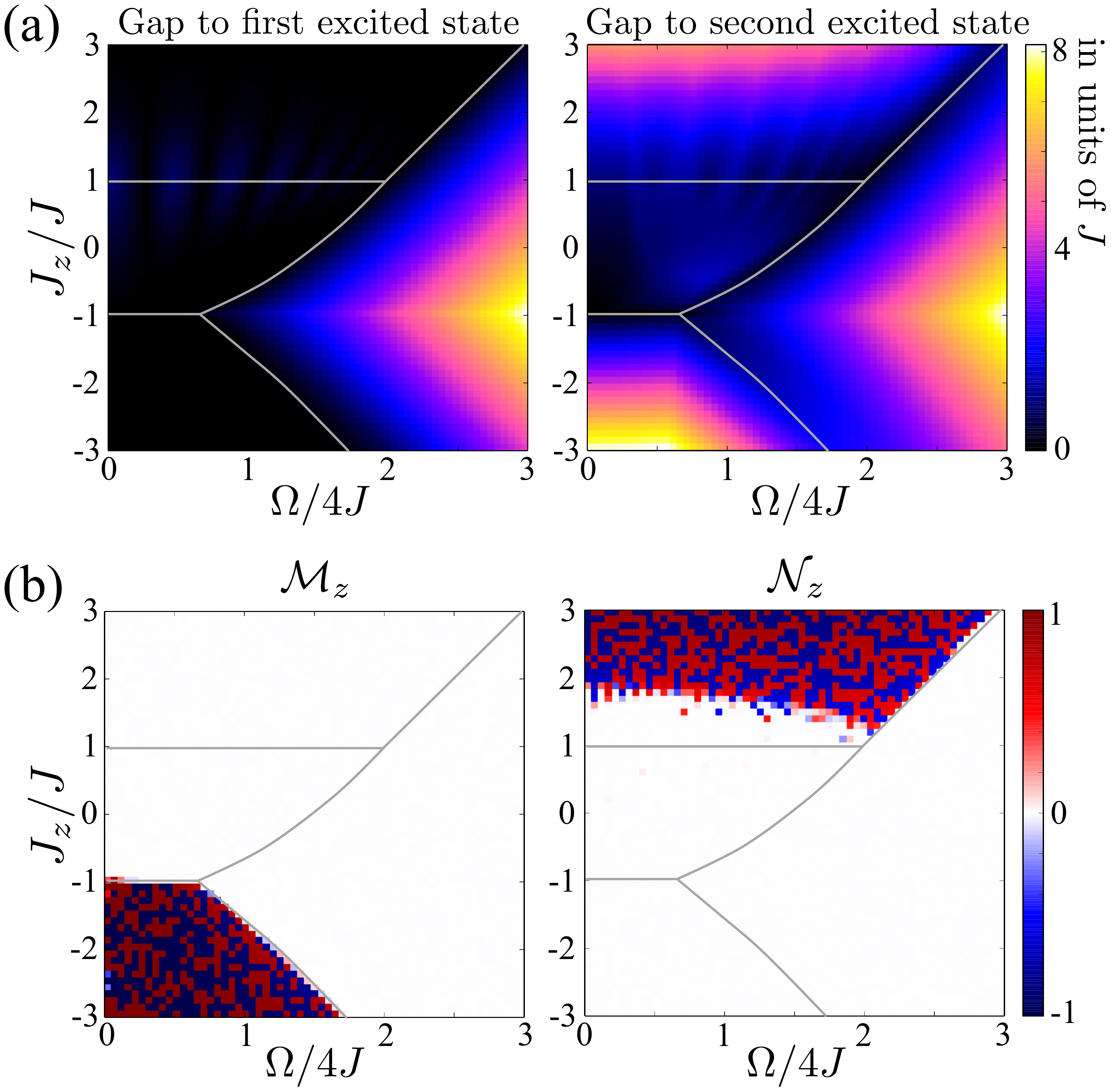}
\caption{(a): Gaps to first and second eigenstates for a chain with periodic boundary conditions and $N=18$ spins. (b): Ferro ($\mathcal{M}_z$) and antiferro ($\mathcal{N}_z = \frac 1 N\sum_j(-1)^j\moy{\sigma_j^z}$) order parameters from $N=90$ MPS calculations.}
\label{fig:gaps}
\end{figure}

Numerical simulations of the spin Hamiltonian are conducted using the ED (Exact Diagonalization) and MPS (Matrix Product States) techniques.

ED\modified{, in which wave-functions and operators are represented exactly,} is used mostly on small systems.
It is used to compute the excitation gaps to the first and second excited states with periodic boundary conditions, shown on Fig.~\ref{fig:gaps}.
For $\mathbb{Z}_2$ symmetry breaking phases, the gap to the first excited state must vanish in the thermodynamical limit, while the gap to the second eigenstate must vanish only on critical lines.
This expected behavior is qualitatively well reproduced numerically in spite of finite-size effects around the $J_z=J$ line, reminiscent of the magnetization plateaus.
\modified{In the time evolution calculation with ED, which includes the atomic motion, we first discretize the continuous time evolution of the Hamiltonian into a staircase function with small steps. On each step, we evolve the wave-function using the expansion of the exponential (well-suited for sparse matrices) and use the criteria of unitary evolution at machine precision to truncate the expansion. }

The MPS calculations are performed using the ITensor library \cite{ITensor}. We use typically  up to 1200 kept states for $N=90$ spins with open boundary conditions.
In many regions of the phase diagram, there are almost classical low-lying excited states, in which the algorithm gets easily trapped, even on small systems.
To help circumvent this issue, we include noise in the reduced density-matrix \cite{White2005} for the first sweeps of the algorithm. 
Furthermore, deep in the symmetry broken phase, the ground state is almost degenerate on large systems (all eigenstates are eigenvectors of the $\mathbb{Z}_2$ symmetries). The MPS algorithm converges thus towards a superposition of these finite-size ground states that effectively breaks the $\mathbb{Z}_2$ symmetries, and that have a lower entanglement entropy. This is illustrated on Fig.~\ref{fig:gaps}, where the local ferromagnetic and N\'eel order parameters are computed from local magnetization. The algorithm randomly converges towards one of the two symmetry breaking states.  The N\'eel order along $y$ never shows up on local observables simply because the algorithm works with real states (the Hamiltonian is purely real).

\section{Loss mechanisms}
\label{app:losses}

The spontaneous emission rate inhibition results from the reduction of the classical electromagnetic field mode density at the atomic emission frequency. It can thus be computed with a classical approach~\cite{ENS_HOUCHES90,TXT_HINDS}. For an atom in the middle of an ideal, infinite plane parallel capacitor (plate separation $D$ along the $OZ$ axis), the spontaneous emission rate modification factors $C_\sigma$ and $C_\pi$ for $\sigma$- and $\pi$-polarized transitions (w.r.t. $OZ$) at wavelength $\lambda$ respectively read
\begin{eqnarray}
C_\sigma&=&\sum_{n=1}^{[2D/\lambda]}\,\frac{3\lambda}{4D}\left[ 1+\left(\frac{n\lambda}{2D}  \right) ^2 \right]\sin^2\left( \frac{n\pi}{2} \right)\\
C_\pi&=&\frac{3\lambda}{4D}+\!\!\sum_{n=1}^{[2D/\lambda]}\,\frac{3\lambda}{2D}\left[ 1-\left(\frac{n\lambda}{2D}  \right) ^2 \right]
\cos^2\left( \frac{n\pi}{2} \right)\nonumber
\label{eq:sponmod}
\end{eqnarray}
where the square brackets in the summation limits stand for the integer part. For $D<\lambda/2$, $C_\sigma=0$. The inhibition is perfect in a capacitor below cut-off and a polarization parallel to the plates.

In order to get a more realistic value, we use a numerical approach taking into account the finite size and conductivity of the capacitor. We compare the total power radiated in free space by a $\sigma$-polarized tiny antenna at the 61.41~GHz frequency of the $\ket{48C}\rightarrow \ket{47C}$ transition to that radiated by the same antenna placed in the capacitor. This computation is performed using the CST Microwave Studio software suite. We have first tested the method with a very large capacitor made up of an ideal conductor. The results are in excellent agreement with the predictions of Eqs.~(\ref{eq:sponmod}). We have then computed the spontaneous emission in a finite capacitor with electrodes made of gold cooled below 1~K (conductivity $4.55\,10^{9}\ \Omega^{-1}$m$^{-1}$~\cite{TXT_HANDBOOK96}).  Note that superconducting electrodes cannot be used in this context, since they are incompatible with the directing magnetic field $\mathbf{B}$. The results of this calculation are presented on Fig.~\ref{fig:inhib}. We choose the operating point $D=2$~mm and $a=13$~mm, providing a 50~dB inhibition.

The lifetime of isolated circular atoms is also limited by the absorption of $\pi$-polarized residual blackbody photons. The dominant processes are the transitions from $\ket{nC}$ to the elliptical states $\ket{(n+1)E^\pm}$. Transitions to higher manifolds are negligible, since the matrix elements and the blackbody number of photons per mode drop rapidly with the upper principal quantum number. The capacitor-induced rate enhancement for these transitions (a factor $\approx$1.8) is computed from Eqs.~(\ref{eq:sponmod}). At $T=0.4$~K, a typical base temperature for a $^3$He refrigerator, we find the excitation rates of $\ket{48C}$ and $\ket{50C}$ to be 1/630~s$^{-1}$ and 1/360~s$^{-1}$, respectively. 

For interacting atoms, the circular states get mixed with elliptical states, which can emit or absorb $\pi$-polarized or high-frequency photons. These processes are not inhibited by the capacitor. The numerical diagonalization of the full pair Hamiltonian provides the expansion of the coupled states on the spherical basis. Using these results, we compute the total decay rate of the coupled levels, including spontaneous decay and blackbody-induced transfers modified by the capacitor.

\begin{figure}
\begin{center}
\includegraphics[width=7.5cm]{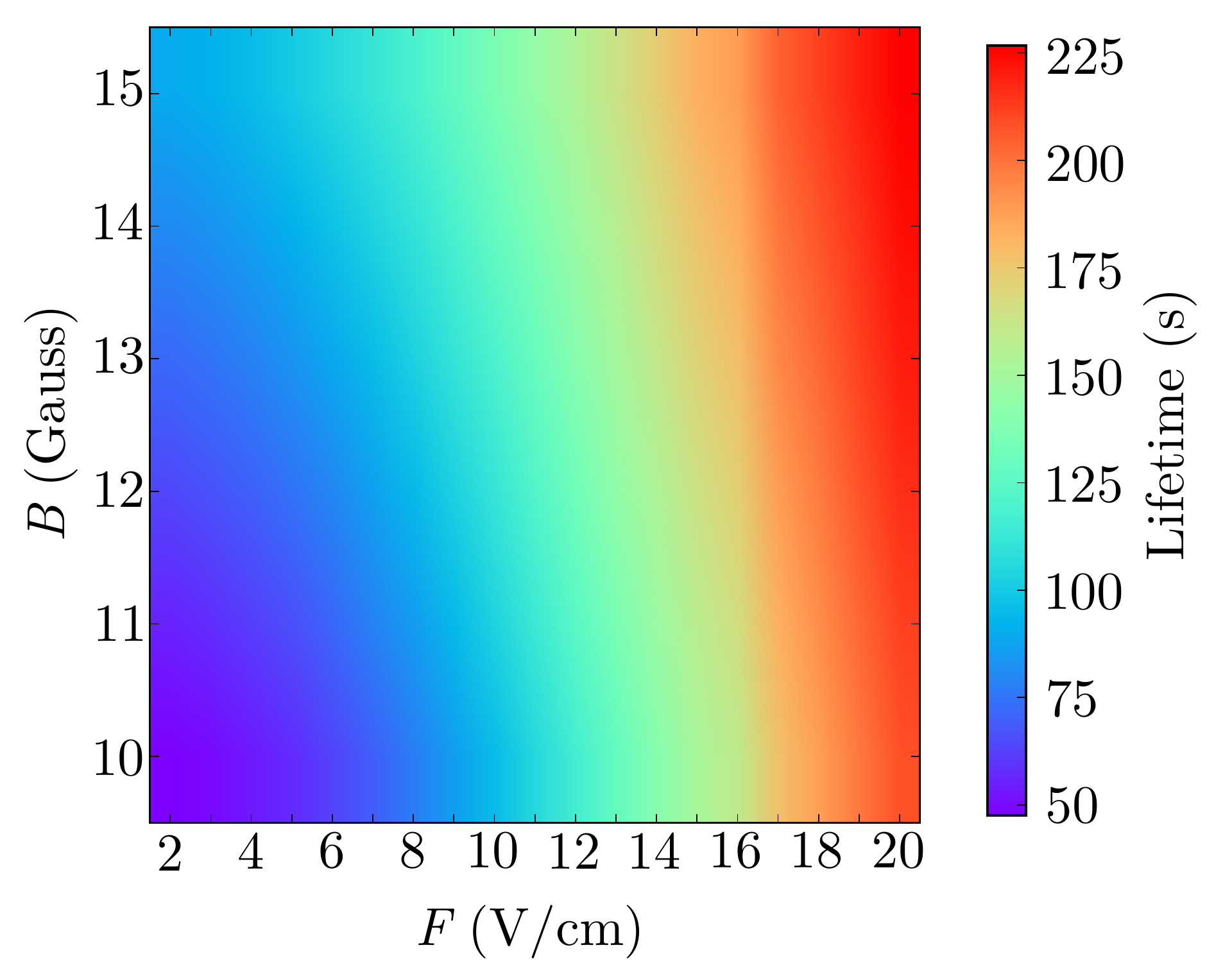}
\end{center} 
\caption{(a) Lifetime of an atom in $\ket{50C}$ interacting with another atom in the same state at a $d=5\ \mu$m distance, as a function of the electric field $F$ and of the magnetic field $B$. }
\label{fig:ratefb}
\end{figure}

Figure~\ref{fig:ratefb} presents a color plot of the  lifetimes, computed in an ideal capacitor, of two $\ket{50C}$ atoms at a $d=5\ \mu$m distance, as a function of the electric field $F$ and of the magnetic field $B$.  Similar results are found for $\ket{48C}$. The lifetime increases with $F$ and $B$, due to the decrease of the  circular state contamination when the directing fields increase (for an isolated atom, the lifetime depends on $F$, but is found to be nearly independent of $B$ for $F>2$~V/cm). Ideally, we should thus aim for the largest field values. However, the tunability of $J_z$ decreases rapidly when $B$ increases (Fig.~\ref{fig:jzj}). To get a flexible simulator, we are thus limited to $\simeq 13$~Gauss, and hence to an individual atom lifetime between 88~s for $F=6$~V/cm and 145~s for $F=12$~V/cm. Note that $F$ can be raised during the chain preparation and detection phases, making radiative losses negligible during these lengthy procedures.

We have also estimated the dipolar relaxation mechanism~\cite{MX_VERHAARDIPREL96}, involving a transition from a pair of atoms in $\ket{50C}$ towards a pair of atoms in $\ket{50E^-}$. This process releases an energy much larger than the trap depth. The two elliptical atoms would thus escape at a high velocity. The matrix element between the initial trapped state and the final high energy plane wave is very small, making the process negligible.

Microwave superradiance~\cite{ENS_PHYSREPSR} does not contribute to a lifetime reduction. First, spontaneous emission and, hence, superradiance on the two-photon  transition from $\ket{50C}$ to $\ket{48C}$ is totally negligible. Superradiance on the one-photon transitions towards the $\ket{49C}$ or $\ket{47C}$ states could be a concern. However, all atoms are in the upper state of the transition. We thus consider only the emission of the first photon in a superradiant cascade, which occurs at a rate  $N$ times larger than for a single atom, a trivial statistical factor. We have already taken into account this effect  when stating that the useful chain lifetime is $1/N$ times that of an individual atom. 

Collisions with the background gas also limit the lifetime. The state-changing cross-sections for the $\ket{20C}$ circular state colliding with Helium gas at room temperature have been calculated  for quite a few final states in~\cite{MX_PRUNELECIRCCOLL85}. Comparable estimates are given in~\cite{MX_MATSUZAWACIRCCOLL84}. Reference~\cite{MX_DEPRUNELECIRCCOLL86} shows that these cross-sections are nearly independent of the electric field, up to 0.2 times the ionization threshold (i.e. up to 20~V/cm for $n=50$). 

Extrapolating to all final states the cross-sections given in~\cite{MX_PRUNELECIRCCOLL85}, we estimate the total cross-section $\sigma_c$ to be of the order of 2000 atomic units for $\ket{20C}$. Intuitively, it should scale as the surface of the circular orbital, a torus with main radius $a_0n^2$ and minor radius $a_0n^2/\sqrt n$. We infer an order of magnitude estimate $\sigma_c\simeq50\,000\,a_0^2$ for $\ket{50C}$, about 10 times the geometric cross-section. The collision lifetime is thus 400~s at a gas density $2.\,10^{11}$~m$^{-3}$, corresponding to $2.6\times10^{-14}$~mbar at 1~K. Such vacuum conditions can be met easily in a cryogenic environment~\cite{MX_GABRIELSEANTIPROTON90,ION_WERTHCRYOION98}, due to the intense cryopumping by all surfaces around the atoms.

Laser trapping competes with photoionization. For low angular momentum Rydberg states, photoionization is fast, with a lifetime in the $\mu$s range for realistic traps~\cite{MX_SAFFMANRYDLOGICDETAILS05}. The situation is radically different for the circular levels~\cite{MX_RAITHELTRAP00}. They are nearly impervious to photoionization. In simple terms, the Rydberg electron absorbs an optical photon with a high momentum only when coming  close to the core, a situation which never happens for circular states. 

The hydrogenic photoionization cross-section $\sigma_\omega(n,\ell)$ at frequency $\omega$ for the state $\ket{n,\ell}$ is computed for isotropic and unpolarized radiation in~\cite{MX_BEZOUGLOVBBION07}. It can be used for an order of magnitude estimate in a polarized laser beam:
\begin{equation}
\sigma_\omega(n,\ell)=\frac{4\ell^4}{9cn^3\omega}\left[ K_{2/3}^2\left( \frac{\omega\ell^3}{3} \right)+ K_{1/3}^2\left( \frac{\omega\ell^3}{3} \right)\right]\ ,
\label{eq:sigmaion}
\end{equation}
where all quantities are expressed in atomic units and where $K_\nu(x)$ is the modified Bessel function of the second kind. For large $\ell$ values and a laser field at a $1\ \mu$m wavelength, the argument of the Bessel functions is large (170 for $\ket{50C}$). We can thus use the asymptotic expansion of $K_\nu(x)$ to lowest order. We get, in SI units:
\begin{equation}
\sigma_\omega(n,\ell)=a_0^2\frac{4\pi}{3}\frac{\alpha\ell}{\omega' n^3}e^{-2\omega' \ell^3/3}\ ,
\end{equation}
with $\omega'=\omega/(2 R_y/\hbar)$, $R_y$ and $\alpha$ being, respectively, the Rydberg and fine structure constants. The cross-section decreases exponentially with $\ell$, down to about $10^{-175}$~m$^{-2}$ for $\ket{50C}$.  A simple estimate based on the wavefunctions in $P$-representation confirms this order of magnitude. Note also that the photoionization rates have been measured as a function of $\ell$ up to $\ell=7$~\cite{MX_SAFFMANRYDLOGICDETAILS05}. The exponential decrease with $\ell$ is conspicuous on these data. The extrapolation to the circular states confirms that photoionization is  indeed negligible.

Another loss channel is the elastic diffusion of the trapping laser by the nearly-free Rydberg electron. This Compton-like process is different from photoionization. The electron receives a momentum kick corresponding to a rather large recoil energy (300~MHz), of the order of the Stark levels separation [$\approx 100$ MHz/(V/cm)]. A diffusion may thus cause a transition towards an elliptical state.  The diffusion cross-section can be evaluated with the classical Thompson diffusion model. Averaging the laser intensity on the atomic motion in the actual trap (peak-to-peak amplitude $\simeq$ 70~nm) and on the electronic motion around the core ($r_{50}=125$~nm), we find that the average time between diffusions is 180~s. This is a worst case estimate of the contribution to the circular state lifetime, since not all diffusion events are expected to change the atomic state.

\begin{table}
\begin{tabular}{|l|c|}
\hline
Cause & Lifetime (s)\\
\hline
\hline
Residual spontaneous emission & 2500\\
\hline
Blackbody induced processes & 630\\
\hline
Level mixing    &   88\\
\hline
Dipolar relaxation &$\infty$ \\
\hline
Photoionization & $\infty$\\
\hline
Collisions with background gas at $10^{-14}$ torr & 400\\
\hline
Compton elastic diffusion in trap & $> 180$ \\
\hline
\hline
\textbf{Predicted lifetime} & \textbf{47}\\
\hline
\end{tabular}
\caption{Decay channels for a pair  $48C$ atoms at $d=5\ \mu$m, $F$=6 V/cm and $B$=13 Gaus. The corresponding lifetimes  are given for a single atom in seconds.}
\label{tab:decay}
\end{table}

Adding all relevant sources of losses, summarized in Table~\ref{tab:decay}, we find an individual atomic lifetime of 47~s, leading to a 1.2~s lifetime for a 40-atom chain. 

\section{Ponderomotive trap}
\label{app:trap}

The trap is formed by the combination of a  standing wave produced by the interference at a small angle between two elongated Gaussian 1$\ \mu$m-wavelength laser beams  (1.45~W each for $d=5\ \mu$m and 2.8~W each for $d=7\ \mu$m) together with a 1$\ \mu$m-wavelength Laguerre-Gauss beam of order $\ell=1;\ p=0$ and waist $w_0=7\ \mu$m (0.5~W power). The intensity of the Laguerre-Gauss beam at a distance $r$ from the $OX$ symmetry axis  in its focal plane  reads
\begin{equation}
I(r)\propto\left(\frac{r\sqrt 2}{w_0}\right)^2 e^{-2 r^2/w_0^2}\ ,
\end{equation}
providing a quadratic trapping potential for small motion. The total depth of the transverse trap is then 6~MHz (300~$\mu$K), while that of the longitudinal lattice is 4~MHz (200~$\mu$K). Near the trap center, the ponderomotive potential is harmonic with trap frequencies $\omega_Y/2\pi=\omega_Z/2\pi=12$~kHz and $\omega_X/2\pi=24$~kHz.

The ponderomotive potentials estimated above assume that the electron has a fixed position in the trap. In fact, it orbits around the core. As shown in~\cite{MX_RAITHELTRAP00}, the ponderomotive energy must be averaged over the electron probability density in the circular state $\ket{nC}$. This average can of course be performed numerically. 

An excellent analytical approximation is obtained by assuming that the electron is on the Bohr orbit with radius $r_n$, in the $XOY$ plane. Using the harmonic approximation to the ponderomotive potential, it is easy to show that the averaging results in a simple offset  on the trapping potential, $M(\omega_X^2+\omega_Y^2) r_n^2/4$, where $\omega_X$ and $\omega_X$ are the trap frequencies for a motion in the plane of the circular orbit. This offset amounts to $h\times22$~kHz for $n=50$. We have checked that this simple model differs from the numerical integration over the electron's probability density by less than 4\%.

Such an offset does not change the trap characteristics. The offsets experienced by $\ket{50C}$ and $\ket{48C}$ differ by 1.7~kHz, resulting in a constant shift of the atomic transition frequency. We thus expect that, to first order, the motion of the atoms in the trap does not contribute to any dephasing of the spin states.

In order to estimate the residual motional dephasing, we must include the anharmonicity of the trapping potential. The dominant effect corresponds to the motion along $OX$. Using the numerical potential averaging, we find that the atomic transition frequency varies quadratically with $X$, being shifted by 12~Hz for $X=70$~nm (this shift can be interpreted as a $\simeq 10^{-3}$ relative difference in the trapping frequencies for the two levels). For a motion in the trap with a 65~nm amplitude (prediction of the numerical simulations of the evaporation process for small chains), this corresponds to a $\approx$160~ms coherence lifetime, much larger than the spin exchange time $\tau_{ex}$.

The flip-flops of the spins in the chain evolution slightly change the interatomic van der Waals forces and thus the equilibrium atomic positions. If this modification was large, this would lead to an entanglement between the spin-chain dynamics and its motional excitations (phonons). This would be  a rich and complex situation, the exploration of which is an interesting perspective for the future of this simulator~\cite{MX_KIMBLESPINCHAIN17}. Nevertheless, we first aim at minimizing this effect and thus choose a tight enough trap.

It is easy to estimate an order of magnitude of the atomic displacement from the center of the trap, $\beta$, in units of the ground-state extension $\Delta X_0$
\begin{equation}
\beta=\frac{4\pi J\eta}{\Omega_X}\ ,
\end{equation}
where $\eta=6\Delta X_0/d$ plays  the role of the Lamb-Dicke parameter of ion traps. Here, for $d=5\ \mu$m, $\eta=6.4\times\,10^{-2}$ and $\beta=0.1$. The atomic displacement being much smaller than the ground-state extension, the entanglement with the motion is negligible. We indeed predict from an explicit analytical model in the simple case of two atoms that the exchange between the spins is not appreciably modified.  Note that the situation would be much worse when using a dipole-allowed transition to encode the spins. For instance, for the $49C-50C$ one-photon transition, the exchange coupling is of the order of 12~MHz. The resulting forces are strong enough to expel the atoms from the trap!

\section{Evaporation process}
\label{app:evap}

We have performed a detailed simulation of the deterministic preparation of a $N \lesssim 40 $ atom chain. We start from a thermal cloud cooled near quantum degeneracy in an elongated dipole trap formed by a 780~nm-wavelength focused laser beam, displaced adiabatically from the atom chip to the science capacitor. We assume a state-of-the-art~\cite{MX_ARIMONDORYDLATT11} cloud of about 2000 atoms with a $\sim 1$ mm length.

We turn off the dipole trap and apply a $1\ \mu$s-long laser pulse to bring the atoms into the $50S$ Rydberg state in the dipole blockade regime. We use 780~nm- and 480~nm-wavelength lasers~\cite{ENS_CHIPSPECTRO14}, tuned on resonance with the two-photon transition from $5S$ to $50S$, and away from resonance with the intermediate $5P$ state. The final positions of the excited Rydberg atoms are simulated using a Monte-Carlo rate equation model including the laser line-width (250~kHz)  and the van der Waals interactions~\cite{ENS_RYDEXCITEXX}. About 100 Rydberg atoms are excited, separated by $9\pm3\ \mu$m.  For such separations, the van der Waals interaction between the atoms is weak, comparable to the laser line-width. 

This excitation stage is immediately followed by the transfer into the circular state in an adiabatic rapid passage sequence, lasting a few microseconds~\cite{ENS_QZDEXP14}. The transfer is induced by a $\sigma^+$-polarized radio-frequency field produced by the four electrodes on the side of $S$.  We finally apply a short pulse of a resonant $780$~nm  laser to push out the remaining ground-state atoms. The motion of the atoms is negligible during the preparation stage, lasting $\approx 10\ \mu$s. The final atomic velocities  are randomly chosen, with a thermal distribution at a 1$\ \mu$K temperature.

\begin{figure}
\begin{center}
\includegraphics[width=7.5cm]{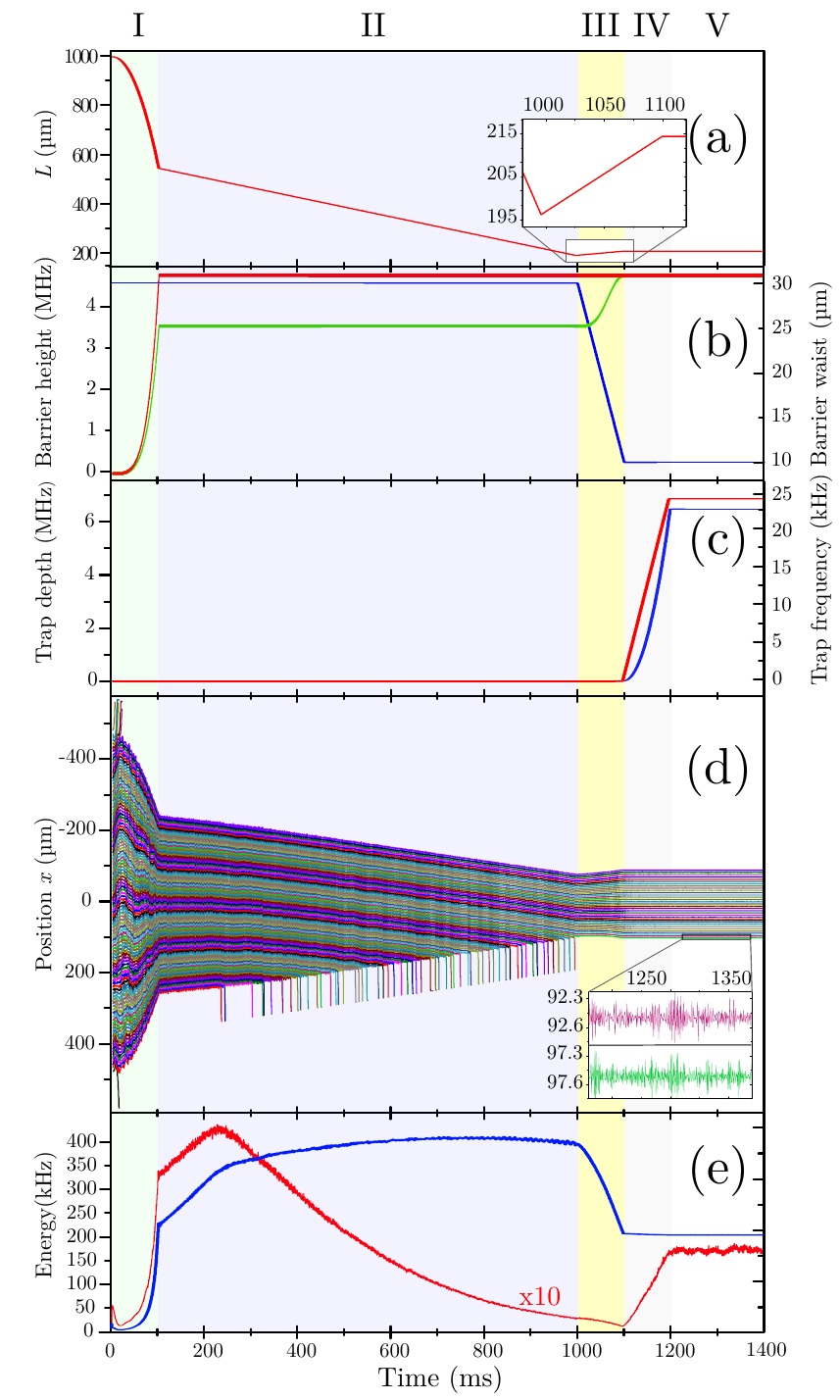}
\end{center} 
\caption{Deterministic chain preparation sequence as a function of time (in milliseconds). (a) distance $L$ between the plug beams. (b) height of the plug beam barriers (red and green lines for the  right and left plugs respectively) and plug beam waists (black line). (c) Longitudinal trap depth (solid line) and oscillation frequency (dashed line). (d) atomic trajectories. The trajectories of ejected atoms are interrupted after a short time for clarity. The inset exhibits the small final residual longitudinal motion. (e) Atomic kinetic energy (red line) and van der Waals potential energy (blue line) averaged over 100 realization of the evaporation sequence. }
\label{fig:sequence}
\end{figure}

Figure~\ref{fig:sequence}(a-c) presents the timing (total duration 1.3~s) of the optimized chain preparation sequence for $N=40$ starting from this initial configuration as well as the results of  a numerical simulation of the 1-D atomic trajectories (we have also performed some 3-D simulations to estimate the transverse atomic motion). This sequence  should be performed with the largest possible $F$ and $B$ values to limit the radiative losses (Appendix~\ref{app:losses}). The sequence is divided in four successive phases:

\begin{enumerate}[label=\Roman* --]

\item Switch-on of  the Laguerre-Gauss transverse trap,  in combination with two `plug' beams (100~ms).  The $1\ \mu$m-wavelength Gaussian plug beams have a $30\ \mu$m waist (this large value results in  a smoother evaporation in phase II). They are initially separated by $L=1$~mm. The height of the associated barriers is smoothly raised from zero to 4~MHz (left beam) or 3~MHz (right beam) -- panel (b). We simultaneously quickly compress the chain by reducing the distance $L$ from 1~mm down to 0.5~mm -- panel (a). This fast compression saves time without significantly modifying the preparation efficiency.

\item Actual evaporation until the required atom number is reached (1000~ms).  The distance $L$ between the two plug beams is slowly reduced. The atomic chain is compressed, building up the repulsive van der Waals forces. The last atom on the weak plug side is expelled out of the trap as soon as its energy exceeds the height of the barrier. This phase stops here at $L=208\ \mu$m to reach the target value $N=40$.

\item Final adjustment of the chain (100~ms). The weak plug barrier is raised to 4~MHz, preventing further evaporation. In the meantime the waists of the plug beams are reduced -- panel (b) -- to provide a finer control of the atomic positions (this stage, experimentally complex, could be replaced by the adiabatic switching-off of the 30 $\mu$m-waist plug beams and the simultaneous adiabatic switching-on of 10 $\mu$m-waist beams). The length $L$ is slightly adjusted [inset in Fig.~\ref{fig:sequence}(a)] to provide a final $d=5\ \mu$m interatomic distance

\item Adiabatic installation of the longitudinal lattice (100~ms) -- panel (c). The amplitude of the residual motion in the traps is accordingly reduced.
\end{enumerate}

The 1-D classical dynamics simulation is complex, since the motion of these coupled atoms is chaotic. The exponential sensitivity to the initial conditions makes it necessary to compute statistics over many realizations. Many numerical methods do not conserve the total energy, resulting in artificial excitation or damping of the system. We thus use a symplectic integrator with a sixth-order Runge-Kutta-Nystr\"om method~\cite{MX_BLANESINTEG02}.

Figure~\ref{fig:sequence}(d) presents the atomic trajectories in one of these simulations. The four phases are clearly apparent. In the first one, during the installation of the plug beams and the fast compression, rapid atomic escapes occurs from both sides while the plugs are still weak. This initial evaporation stops after $\approx 100$~ms. The chain is then compressed more slowly. Evaporation above the weak plug resumes at the beginning of phase II, atoms escaping in the positive $OX$ direction. The evaporation events seem to reduce the residual motion of the remaining atoms. 

This qualitative insight is confirmed in Fig.~\ref{fig:sequence}(e), which presents the  kinetic and potential van der Waals energies per atom averaged over 100 realizations of the evaporation sequence. During the evaporation stage II, the kinetic energy clearly decreases. The evaporation above the plug barriers provides a cooling reminiscent of the evaporative cooling~\cite{MX_KLEPPNEREVAP88}. The final motion of the 40-atom chain after stage IV has a typical extension $\Delta X=110$~nm (see inset in Fig.~\ref{fig:sequence}(e)). The 3-D simulations indicate that the transverse motion extensions $\Delta Y$ and $\Delta Z$ are of the same order of magnitude as $\Delta X$. Note that the transverse motion does not appreciably modify the interatomic distance.

Running 100 times the simulation, continued for 1.4~s to the end of the evaporation stage II, when the final chain only contains one atom, we obtain the average number of atoms and its standard deviation as a function of the final length $L$ presented in Fig.~\ref{fig:nevap}. 

During evaporation, the atoms, particularly those close to the end of the final chain, transiently experience rather large trap laser intensities. We have estimated the associated loss rate due to Compton diffusion events in a full 3-D simulation. It is small, less than 3\% for the atoms at the extremities of the chain, about 1\% for the bulk atoms. Selective microwave transitions from the circular states towards a lower manifold and field-ionization of the remaining atoms could be used for a final purification of the chain before switching on the longitudinal lattice.

The atomic detection stage simply resumes the evaporation stage II after removing the longitudinal lattice and lowering the right plug beam. This process is clearly less critical, the only requirement being to keep the order of the atoms. The velocity of the ejected atoms in the guiding LG beam is determined by the height of the weak plug, $0.16$~m/s for 3~MHz.  The atoms thus reach the detection region, about 2~cm away, after a 125~ms delay, short as compared to their lifetime. 

\bibliographystyle{prx_apsrev4-1}
%

\end{document}